\documentclass[12pt]{article}

\usepackage{amstext,amsmath,amssymb,amsfonts,bbm}
\usepackage[latin1]{inputenc}
\usepackage{epsfig}
\usepackage{dsfont}
\usepackage{hyperref}
\usepackage{amsthm}
\usepackage{subfigure}
\usepackage{color}
\usepackage{multirow}
\usepackage{psfrag}
\usepackage{graphicx}
\usepackage{extarrows}
\usepackage{authblk}

\usepackage{braket}
\usepackage{mathtools} 
\usepackage{bbold}
\usepackage{stmaryrd} 
\usepackage{color}
\usepackage{geometry}

\usepackage{tikz-feynman}

\theoremstyle{plain}

\theoremstyle{definition}

\begin{document}
 
\title{Non-Gaussian disorder average in the Sachdev-Ye-Kitaev model}
\author[1]{T. Krajewski\thanks{thomas.krajewski@cpt.univ-mrs.fr}
}
\author[2,3]{M. Laudonio \thanks{matteo.laudonio@u-bordeaux.fr}}

\author[2,4]{R. Pascalie\thanks{romain.pascalie@u-bordeaux.fr}}

\author[2,5]{ A. Tanasa\thanks{ntanasa@u-bordeaux.fr}}

\affil[1]{\small Aix Marseille Univ, Universit\'e de Toulon, CNRS, CPT, Marseille, France\\}

\affil[2]{\small Universit\'e de Bordeaux, LaBRI, CNRS UMR 5800,  Talence, France, EU\\}

\affil[3]{\small Department of Applied Mathematics, University of Waterloo, Waterloo, Ontario, Canada}

\affil[4] {\small Mathematisches Institut der Westfalischen Wilhelms-Universitaat,  M\"unster, Germany, EU}

\affil[5]{\small 
H. Hulubei Nat. Inst.  Phys.  Nucl. Engineering, Magurele, Romania, EU\\

IUF Paris, France, EU}

\maketitle
\abstract{We study the effect of non-Gaussian average over the random couplings in a complex version of the celebrated Sachdev-Ye-Kitaev (SYK) model. 
Using a Polchinski-like equation and random tensor Gaussian universality, we show that the effect of this non-Gaussian averaging  leads to a modification of the variance of the Gaussian distribution of couplings  at leading order in $N$. We then derive the form of the effective action to all orders. An explicit computation of the modification of the variance in the case of a quartic perturbation is performed for both the complex SYK model mentioned above and the SYK generalization proposed in D. Gross and V. Rosenhaus, JHEP  1702 (2017) 093.}

\newpage

\section{Introduction}

The Sachdev-Ye model \cite{sachdev} has been initially proposed and intensively studied in a condensed matter setting. In a series of talks \cite{kitaev}, Kitaev introduced a simplified  version of this model and showed
it can be a particularly interesting toy-model for AdS/CFT physics. The Sachdev-Ye-Kitaev (SYK) model has ever since attracted a huge amount of interest for both condensed matter and high energy physics - see, for example, \cite{georges}, \cite{maldacena}, \cite{polchinski}, \cite{Gross},
\cite{Bonzom:2017pqs}, \cite{Bonzom:2018jfo}, \cite{Carrozza:2018psc}, 
or the review articles \cite{Sarosi} or \cite{Rosenhaus}.

More specifically, the SYK model is a quantum-mechanical model with $N$ fermions with random interactions involving $q$ of these fermions at a time. Each coupling $J$ is a variable drawn from a random Gaussian distribution. In this paper we investigate the behaviour of the model when this Gaussianity condition is relaxed. 

We first work with a version of the SYK model containing $q$ flavors of complex fermions, each of them appearing once in the interaction. This model is very close in the spirit to the colored tensor model (see the book \cite{gurau-book}) and it is a particular case of a complex version of the Gross-Rosenhaus SYK generalization proposed in \cite{Gross}. This particular version of the SYK model has already been studied in \cite{complete}, \cite{quenched}, \cite{DartoisCFT}, \cite{Bonzom:2017pqs} and \cite{Fusy:2018xax}.

Following the approach proposed in \cite{Krajewski} for tensor models and group field theory (see also \cite{Krajewski2}, \cite{Krajewski3} and \cite{Krajewski4}),  
we first use a Polchinski-like flow equation to obtain Gaussian universality. This Gaussian universality result for the colored tensor model was initially proved in \cite{universality}.
Let us also mention here that this universality result for colored tensor models was also exploited in \cite{Bonzom}, in a condensed matter physics setting, 
to identify an infinite universality class of infinite-range $p-$spin glasses with non-Gaussian correlated quenched distributions.

In this paper we further obtain the effective action for the non-Gaussian averaged complex SYK model studied here.  We show that the effect of this non-Gaussian averaging is a modification of the variance of the Gaussian distribution of couplings at leading order in $N$.

We then choose a specific quartic perturbation (known in the tensor model literature as a pillow or  a melonic quartic perturbation, see for example, \cite{Sanchez}, \cite{Pascalie}, \cite{Carrozza} or \cite{Klebanov} or the TASI lectures on large $N$ tensor models \cite{Klebanov2}) and, using the Hubbard-Stratanovitch (or the intermediate filed representation) for the disorder $J$, we explicitly compute the first order correction of the effective action and the modification of the Gaussian distribution of the couplings $J$ at leading order in $N$. We then generalize these explicit calculations for the  Gross-Rosenhaus SYK model proposed in \cite{Gross} (the fermionic fields being this time real) and, as above, we obtain  the first order correction of the Gross-Rosenhaus SYK effective action and the modification of the Gaussian distribution of the couplings $J$ at leading order in $N$.

\medskip

For the sake of completeness, let us mention that in \cite{rusii}, the $4-$point function of SYK model in a double-scaling limit was computed and the random couplings did not necessarly had to be independ and Gaussian - it was enough for these random couplings to be taken  independent random variables, with zero mean and uniformly bounded moments independent of $N$.

\medskip

Our paper is organized as follows. In the following section we introduce the complex SYK model we initially work with and we express the non-Gaussian potential as a sum over particular graphs. In Section $3$, the Gaussian universality result is exhibited, using a Polchinski-like equation. The following section is dedicated to the study of the the effective action for this model. Sections $5$ and resp. $6$ we perform our explicit calculations for quartic perturbations for the complex SYK and resp. the (real) Gross-Rosenhaus SYK generalization. The last section lists some concluding remarks. For the sake of completeness, we add an appendix which constructs
 the Dyson-Schwinger equations
for the intermediate field used in this paper. This construction follows the lines 
of \cite{Nguyen:2014mga}, and it is done for both the complex SYK and resp. the (real) Gross-Rosenhaus SYK generalization studied here.

\section{A complex SYK model with non-Gaussian disorder}

As already announced in the previous section, we study here a complex SYK model with $q$ complex 
fermions $\psi^{a}_{i_{a}}(t)$, where the label $a=1,..,q$ is the flavor and each fermion carries an index $i_{a}=1,...,N$. 
The action writes:
\begin{align}
\label{noi}
S_{J}(\psi, \bar{\psi})=
\int dt \bigg(\sum_{a,i_{a}} \bar{\psi}^{a}_{i_{a}}\partial_{t}\psi^{a}_{i_{a}}
+\text{i}^{\frac q2}\sum_{i_{1},\dots,i_{q}} \bar{J}_{i_{1},\dots,i_{q}}\psi^{1}_{i_{1}}\cdots \psi^{q}_{i_{q}}+
\text{i}^{\frac q2}\sum_{i_{1},\dots,i_{q}}J_{i_{1},\dots,i_{q}} \bar{\psi}^{1}_{i_{1}}\cdots \bar{\psi}^{q}_{i_{q}}\bigg).
\end{align}
Here $J_{i_{1},\dots,i_{q}}$ is a rank $q$ tensor that plays the role of a coupling constant. 
As already mentioned in the Introduction, this model is close in spirit to tensor models and is a particular case of the Gross-Rosenhaus generalization of the SYK model.

For the sake of completeness, let us mention that a bipartite complex SYK-like tensor model 
(without any fermion flavors and) 
with $ O(N)^3$ symetry 
was studied in the TASI lectures \cite{Klebanov2}.
It was then found that one of the operators has a complex scaling dimension, which suggests that the nearly-conformal large $N$ phase of the bipartite model is unstable.

\begin{figure}
    \centering
    \includegraphics[scale=0.95]{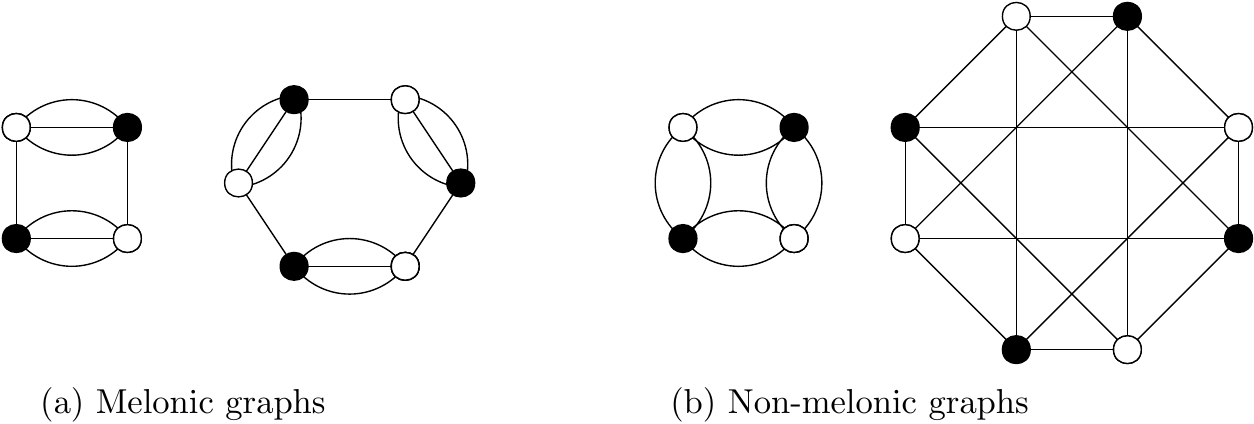}
    \caption{Examples of melonic and non-melonic graphs for $q=4$.}
    \label{fig:bubbles}
\end{figure}

The model \eqref{noi} we study here is subject to quenched disorder - we average  the free energy (or connected correlation functions) over the couplings $J$.
The most convenient way to performe this is through the use of  replicas. We thus add an extra replica index $r=1,\dots,n$ to the fermions. One has: 
\begin{align}
\langle \log Z(J)\rangle_{J}=\lim_{n\rightarrow 0}\frac{\langle Z^{n}(J)\rangle_{J}-1}{n},
\end{align}
with
\begin{align}
Z^{n}(J)=\int\prod_{1\leq r\leq n}[d\psi_{r}][d \bar{\psi}_{r}]\exp\sum_{r}S_{J}(\psi_{r}, \bar{\psi}_{r}).
\end{align}
The averaging over $J$ is performed with a possibly non-Gaussian weight:
\begin{align}
\langle Z^{n}(J)\rangle_{J}=\frac{\int dJ d \bar{J}\, Z^{n}(J)\exp \big[-\big[\frac{N^{q-1}}{\sigma^{2}}J \bar{J}+V_{N}(J, \bar{J})\big]\big]}{\int dJd \bar{J} \exp\big[-\big[\frac{N^{q-1}}{\sigma^{2}}J \bar{J}+V_{N}(J, \bar{J})\big]\big]}.
\end{align}
We further impose that the potential $V_N$ is invariant under independent unitary transformations: 
\begin{align}
J_{i_{1},\dots,i_{q}}\rightarrow \sum_{j_{1},\dots,j_{q}}U^{1}_{i_{1}j_{1}}\cdots U^{q}_{i_{q}j_{q}}J_{j_{1},\dots,j_{q}, }\qquad
 \bar{J}_{i_{1},\dots,i_{q}}\rightarrow \sum_{j_{1},\dots,j_{q}} \bar{U}^{1}_{i_{1}j_{1}}\cdots  \bar{U}^{q}_{i_{q}j_{q}} \bar{J}_{j_{1},\dots,j_{q}}.\label{unitary}
\end{align}
Assuming that the potential $V_N$ is a polynomial (or an analytic function) in the couplings $J$ and $ \bar{J}$, this invariance imposes that the potential can be expanded over particular graphs, as follows. Let us consider (non necessarily connected) graphs\footnote{These graphs are called bubbles in the tensor model literature (see again the book \cite{gurau-book}.)}
 $\Gamma$ with black and white vertices of valence $q$. The edges of such a graph connect only black to white vertices (we thus have bipartite graphs) and are labelled by a color $a=1,\ldots,q$ in such a way that, at each vertex, the $q$ incident edges carry distinct colors (we thus have edge-colored graphs). Let us mention that each edge color of these graphs $\Gamma$ corresponds to a fermion flavor of the model. 
See Fig. \ref{fig:bubbles} for some examples of such graphs:
melonic graphs on figure (a) and non-melonic graphs on figure (b).
A graph is called melonic if for any vertex $v$, there is another vertex $ \bar{v}$ such that the removal of $v$ and $ \bar{v}$ yields exactly $q$ connected components (including isolated lines). 

The most general form of the 
potential is expanded over these graphs as:
\begin{align}
V_{N}(J, \bar{J})=\sum_{\text{graph $\Gamma$}}\lambda_{\Gamma}\frac{N^{q-k(\Gamma)}}
{\text{Sym($\Gamma$)}}
\langle J, \bar{J}\rangle_{\Gamma},
\end{align}
where we have used the shorthand for the contraction of tensors along the graph $\Gamma$
\begin{align}
\langle J, \bar{J}\rangle_{\Gamma}=\sum_{1\leq i_{v,a},\dots,i_{ \bar{v},a}\leq N}
\prod_{\text{white}\atop\text{vertices }v}J_{i_{v,1},\dots,i_{v,q}} 
\prod_{\text{black}\atop\text{vertices } \bar{v}} \bar{J}_{ \bar{i}_{ \bar{v},1},\dots, \bar{i}_{ \bar{v},q}} 
\prod_{\text{edges }\atop e=(v, \bar{v})}\delta_{i_{v,c(e)},i_{ \bar{v},c(e)}}.\label{graphs}
\end{align}
In this expression, $\lambda_{\Gamma}$ is a real number, $k(\Gamma)$ is the number of connected components of $\Gamma$ and Sym($\Gamma$) its symmetry factor. The contraction of indices means that each white vertex carries a tensor $J$, each black vertex a tensor $ \bar{J}$ and that the indices have to be contracted by identifying two indices on both sides of an edge, the place of the index in the tensor being defined by the color of the edge denoted by $c(e)$. 

The Gaussian term  corresponds to a dipole graph (a white vertex and a black vertex, related by $q$ lines) and reads
\begin{align}
\frac{N^{q-1}}{\sigma^{2}}J \bar{J}=
\frac{N^{q-1}}{\sigma^{2}}\sum_{1\leq i_{1},\dots,i_{q}\leq N}
J_{ i_{1},\dots,i_{q}} \bar{J}_{ i_{1},\dots,i_{q}}
\end{align}
Introducing the pair of complex conjugate tensors $K$ and $ \bar{K}$ defined by
\begin{align}
K_{ i_{1},\dots,i_{q}}=\text{i}^{\frac q2}\sum_{r}\int dt 
\psi^{1}_{i_{1},r}\cdots \psi^{q}_{i_{q};r}
\qquad
 \bar{K}_{ i_{1},\dots,i_{q}}=\text{i}^{\frac q2}\sum_{r}\int dt 
 \bar{\psi}^{1}_{i_{1},r}\cdots  \bar{\psi}^{q}_{i_{q};r},\label{definitionK}
\end{align}
 the averaged partition function reads 
\begin{align}
\label{average}
\langle Z^{n}(J)\rangle_{J}=\frac{\int[d\psi][d \bar{\psi}]
\exp\big[-\int dt \sum_{a,i_{a},r} \bar{\psi}^{a}_{i_{a},r}\partial_{t}\psi^{a}_{i_{a},r}\big]\,\int dJ d \bar{J}
\exp\big[-\big[\frac{N^{q-1}}{\sigma^{2}}J \bar{J}+V_{N}(J, \bar{J})+J \bar{K}+ \bar{J}K\big]
\big]}{\int dJd \bar{J} \exp \big[-\big[\frac{N^{q-1}}{\sigma^{2}}J \bar{J}+V_{N}(J, \bar{J})\big]\big]}.
\end{align}
After a shift of variables in the integral over $J$ and $ \bar{J}$, the integral on $J$ and $\bar J$ in the numerator reads
\begin{align}
\exp\bigg[-\frac{\sigma^{2}}{N^{q-1}}K \bar{K}\bigg]
\int dJ d \bar{J}
\exp-\bigg[\frac{N^{q-1}}{\sigma^{2}}J \bar{J}+V_{N}\Big(J-\frac{\sigma^{2}}{N^{q-1}}K, \bar{J}-\frac{\sigma^{2}}{N^{q-1}} \bar{K}\Big)\bigg].
\end{align}
In order to study the large $N$ limit of the average \eqref{average}, we introduce the background field effective potential, with $L=-\frac{\sigma^{2}}{N^{q-1}}K$ and $ \bar{L}=-\frac{\sigma^{2}}{N^{q-1}} \bar{K}$. One has:
\begin{align}
V_{N}(s,L, \bar{L})=-\log
\int dJ d \bar{J}
\exp-\bigg[\frac{N^{q-1}}{s}J \bar{J}+V_{N}\Big(J+L, \bar{J}+ \bar{L}\Big)\bigg]\quad+N^{q}\log\frac{\pi s}{N^{q-1}}\label{effectivepotential}
\end{align}
In this framework, $s$ is a parameter that interpolates between the integral we have to compute, at $s= \sigma^{2}$ (up to a trivial multiplicative constant) and the potential we started with at $s=0$ (no integration and $J= \bar{J}=0$). 
The inclusion of the constant ensures that the effective potential remains zero when we start with a vanishing potential. 
This comes to:
\begin{align}
\int dJ d \bar{J}
\exp\bigg[-\bigg[\frac{N^{q-1}}{\sigma^{2}}J \bar{J}+V_{N}\Big(J-\frac{\sigma^{2}}{N^{q-1}}K, \bar{J}-\frac{\sigma^{2}}{N^{q-1}} \bar{K}\Big)\bigg]\bigg]\nonumber\\
=
\bigg(\frac{N^{q-1}}{\pi s}\bigg)^{N^{q}}
\exp \bigg[-V_{N}\bigg(s=\sigma^{2},L=-\frac{\sigma^{2}}{N^{q-1}}K, \bar{L}=-\frac{\sigma^{2}}{N^{q-1}} \bar{K}\bigg)\bigg].
\end{align}
In the next section, we will derive the large $N$ behavior of the effective potential using a 
Polchinski-like flow equation.

\section{Gaussian universality}

\begin{figure}
    \centering
    \includegraphics{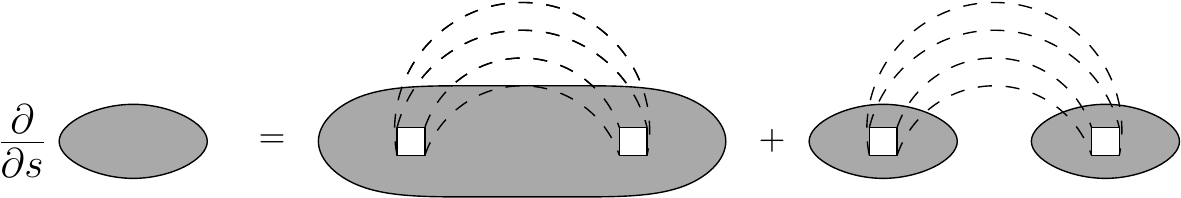}
    \caption{Graphical representation of equation \eqref{ERGE} for $q=4$.}
    \label{fig:polchinski}
\end{figure}

As already mentioned in the Introduction, this section follows the approach proposed in \cite{Krajewski} (see also \cite{Krajewski2}, \cite{Krajewski3} or \cite{Krajewski4}).
Using 
standard QFT manipulations
(see for example, the book \cite{Zinn-Justin}), one can show that the effective potential  $V_{N}(s,L, \bar{L})$ (see eq. \eqref{effectivepotential}) obeys the following differential equation:
\begin{align}
\frac{\partial V}{\partial s}=\frac{1}{N^{q-1}}\sum_{1\leq i_{1},\dots,i_{q}\leq N}\bigg(
\frac{\partial ^{2}V}{\partial L_{i_{1},\dots,i_{q}}\partial  \bar{L}_{i_{1},\dots,i_{q}}}-
\frac{\partial V}{\partial L_{i_{1},\dots,i_{q}}}\frac{\partial V}{\partial  \bar{L}_{i_{1},\dots,i_{q}}}
\bigg)\label{ERGE}
\end{align}
One can represent this equation in a graphical way as shown in Fig. \ref{fig:polchinski}. The first term on the RHS corresponds to an edge closing a loop in the graph and the second term in the RHS corresponds to a bridge (or an 1PR) edge.
This equation is formally a Polchinski-like equation \cite{Polchinski84}, albeit there are no short distance degrees of freedom over which we integrate. In our context it simply describes a partial integration with a weight $s$ and will be used to control the large $N$ limit of the effective potential.

Since the effective potential is also invariant under the unitary transformations defined in eq. \eqref{unitary}, it may also be expanded over graphs as in \eqref{graphs},
\begin{align}
V_{N}(s,L, \bar{L})=
\sum_{\text{graph $\Gamma$}}\lambda_{\Gamma}(s)\frac{N^{q-k(q)}}
{\text{Sym($\Gamma$)}}\langle L, \bar{L}\rangle_{\Gamma},
\label{effectivegraphs}
\end{align}
with $s$ dependent couplings $\lambda_{\Gamma}(s)$. 

Inserting this graphical expansion in the differential equation \eqref{ERGE}, we obtain a system of differential equations for the couplings,
\begin{align}
\frac{d\lambda_{\Gamma}}{ds}=\sum_{\Gamma'/( \bar{v}v)=\Gamma}
N^{k(\Gamma)-k(\Gamma')+e(v, \bar{v})-q+1}\,\lambda_{\Gamma'}-
\sum_{(\Gamma'\cup\Gamma'') /( \bar{v}v)=\Gamma}\lambda_{\Gamma'}\,\lambda_{\Gamma''}
\label{system}
\end{align}
A derivation of the potential $V$ with respect to $L_{i_{1},\dots,i_{q}}$ (resp. $ \bar{L}_{i_{1},\dots,i_{q}}$) removes a white vertex (resp. a black vertex). Then, the summation over the indices  in $i_{1},\dots,i_{q}$ in \eqref{ERGE} reconnects the edges, respecting the
 colors.  

\begin{figure}[ht]
\centering
\includegraphics[scale=0.65]{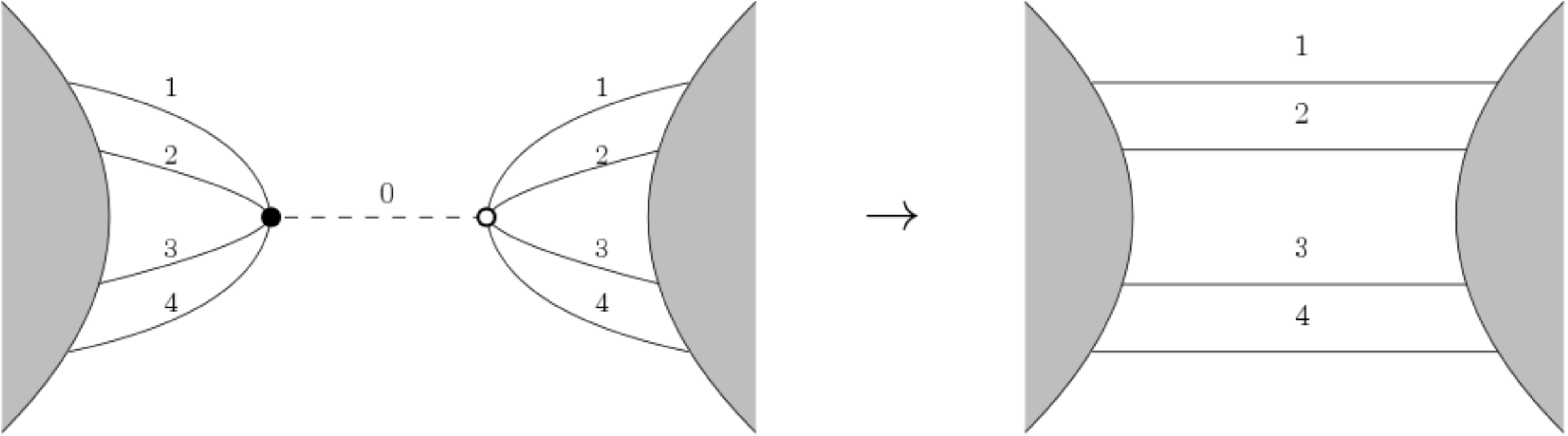}
\caption{Removal of a white and a black vertex and reconnection of the edges.}
\label{remove}
\end{figure}

\begin{figure}[ht]
\centering
\includegraphics[scale=0.65]{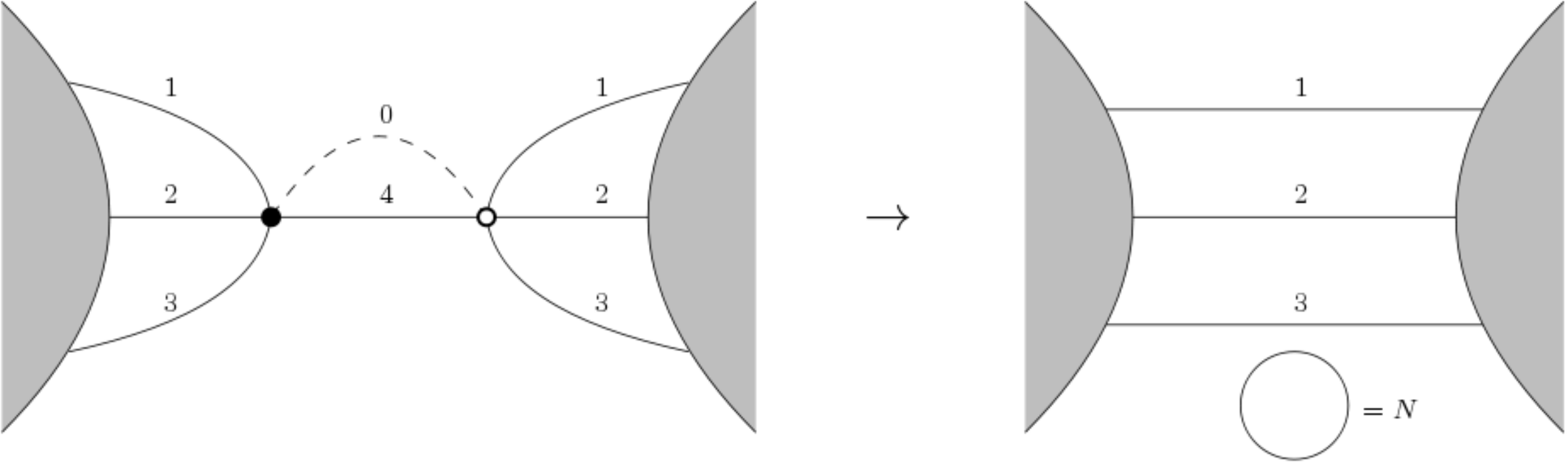}
\caption{Removal of a white and a black vertex and reconnection of the edges creating a loop.}
\label{remove2}
\end{figure}

In the first term on the RHS of \eqref{ERGE}, given a graph $\Gamma$ in the expansion of the LHS, we have to sum over all graphs $\Gamma'$ and pairs of a white vertex $v$ and a black vertex $ \bar{v}$ in $\Gamma'$ such that the graph $\Gamma'/( \bar{v}v)$ obtained after reconnecting the edges (discarding the connected components made of single lines) is equal to $\Gamma$ - see Fig. \ref{remove} and Fig. \ref{remove2}.

The number $e(v, \bar{v})$ is the number of edges directly connecting $v$ and $ \bar{v}$ in $\Gamma$. After summation over the indices, each of these lines yields a power of $N$, which gives the factor  of $N^{e(v, \bar{v})}$. 

The operation of removing two vertices and reconnecting the edges can at most increase the number of connected components (including the graphs made of single closed lines) by $q-1$, so that we always have $k(\Gamma)-k(\Gamma')+e(v, \bar{v})-q+1\leq 0$. We obtain the equality if and only if $\Gamma'$ is a melonic graph. 
Therefore, in the large $N$ limit, only melonic graphs survive in the first term on the RHS of \eqref{system}.

In the second term, we sum over graphs $\Gamma'$ and  white vertices $v\in\Gamma'$ and graphs $\Gamma''$ and black vertices $ \bar{v}\in\Gamma''$, with the condition that the graph obtained after removing the vertices and  reconnecting the lines $(\Gamma'\cup\Gamma'')/( \bar{v}v)$ is equal to $\Gamma$. In that case, the number of connected components necessarily diminishes by $1$, so that all powers of $N$ cancel. 

The crucial point in the system \eqref{system} is that only negative (or null) powers of $N$ appear. It can be written as
\begin{align}
\frac{d\lambda_{\Gamma}}{ds}=
\beta_{0}\big(\left\{\lambda_{\Gamma}\right\}\big)
+\frac{1}{N}\beta_{1}\big(\left\{\lambda_{\Gamma}\right\}\big)
+\dots
\end{align}
As a consequence, if $\lambda_{\Gamma}(s=0)$ is bounded, then $\lambda_{\Gamma}(s)$ is also bounded for all $s$ (i.e. it does not contain positive powers of $N$).

Let us now substitute $L=-\frac{\sigma^{2}}{N^{q-1}}K$ and $ \bar{L}=-\frac{\sigma^{2}}{N^{q-1}} \bar{K}$ in the expansion of the effective potential
\eqref{graphs},
\begin{align}
\label{final}
V_{N}\bigg(s=\sigma^{2},L=-\frac{\sigma^{2}}{N^{q-1}}K, \bar{L}=-\frac{\sigma^{2}}{N^{q-1}} \bar{K}\bigg)=
\sum_{\text{graph $\Gamma$}}\lambda_{\Gamma}(\sigma^{2})\frac{(-\sigma^{2})^{v(\Gamma)}N^{q-k(q)-(q-1)v(\Gamma)}}
{\text{Sym($\Gamma$)}}\langle K, \bar{K}\rangle_{\Gamma}.
\end{align}
Here $v(\Gamma)$ the number of vertices of $\Gamma$. 
The exponent of $N$ can be rewritten as $(q-1)(1-v(\Gamma))+1-k(\Gamma)$. It has it maximal value for $v(\Gamma)=2$ and $k(\Gamma)=1$, which corresponds to the dipole graph. 
This is a reexpression the 
Gaussian universality property of random tensors. 


\section{Effective action}

Taking into account the non-Gaussian quenched disorder, we derive the effective action  for the bilocal invariants,
\begin{align}
{G}_{r,r'}^{a}(t,t')=\frac{1}{N}\sum_{i}\psi^{a}_{i,r}(t_{1}) \bar{\psi}^{a}_{i,r'}(t').\label{bilocal}
\end{align}
Note that these invariants carry one flavour label $a$ and two replica indices $r,r'$. 

To this end, let us come back to the partition function \eqref{average}. We then express the result of the average over $J$ and $ \bar{J}$ as a sum over graphs $\Gamma$ using the expansion of the effective potential \eqref{final} and replacing the tensors $K$ and $ \bar{K}$ in terms of the fermions $\psi$ and $ \bar{\psi}$ (see eq. \eqref{definitionK}).

Each graph $\Gamma$ then involves the combination
\begin{align}
\langle K, \bar{K}\rangle_{\Gamma}=\sum_{1\leq i_{v,a},\dots,i_{ \bar{v},a}\leq N}
&\prod_{\text{white}\atop\text{vertices }v}\sum_{r_{v}}\int dt_{v}
\psi^{1}_{i_{v,1},r_{v}}(t_{v})\cdots\psi^{q}_{i_{v,q},r_{v}}(t_{v})\nonumber \\
&\prod_{\text{black}\atop\text{vertices } \bar{v}}
\sum_{r_{ \bar{v}}}\int dt_{ \bar{v}}
 \bar{\psi}^{1}_{ \bar{i}_{ \bar{v},1},r_{ \bar{v}}}\cdots
 \bar{\psi}^{q}_{ \bar{i}_{ \bar{v},q}, \bar{v}} (t_{ \bar{v}})\prod_{\text{edges }\atop e=(v, \bar{v})}\delta_{i_{v,c(e)},i_{ \bar{v},c(e)}}.
\label{graphs2}.
\end{align}
After introducing the Lagrange multiplier ${\Sigma}$ to enforce the constraint \eqref{bilocal} and assuming a replica symmetric saddle-point, the effective action of our model writes:
\begin{align}
   \frac{\mathcal{S}_{eff}[\mathrm{G},\Sigma]}{N} 
   = &- \sum \limits_{f=1}^{q} \log{\det{\big(\delta(t_1-t_2)\partial_{t}-{\Sigma}_f(t_1,t_2) \big)}}
    +\int \mathrm{d}\mathbf{t} \sum \limits_{f=1}^{4}{\Sigma}_f(\mathbf{t}) {G}_f(\mathbf{t})\nonumber \\
    &-\sum_{\Gamma} N^{-(v(\Gamma)-2)(q/2-1)+1-k(\Gamma)}\mu_{\Gamma}(\sigma^2,\{\lambda_{\Gamma'}\}) \langle {G} \rangle_{\Gamma},
\end{align}

The term $\langle {G} \rangle_{\Gamma}$ associated to a graph $\Gamma$ is constructed as follows:
\begin{figure}
    \centering
    \includegraphics{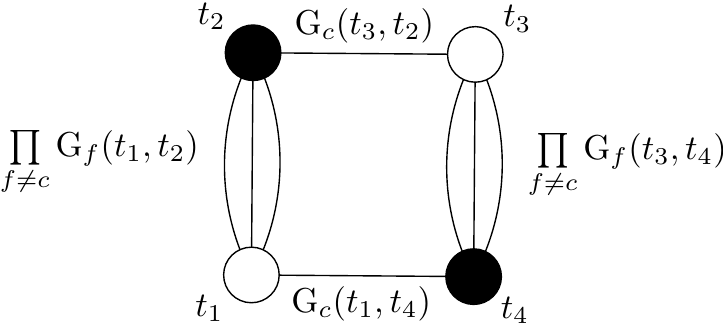}
    \caption{Graphical representation of the term $\langle\mathrm{G}\rangle_{\Gamma}$ for the quartic melonic graph for $q=4$.}
    \label{fig:effpillow}
\end{figure}

\begin{itemize}
\item to each vertex associate a real variable $t_{v}$;
\item to an edge of colour $c$ joining $v$ to $v'$ associate ${G}_{c}(t_{v},t_{v'})$;
\item multiply all edge contributions and integrate over vertex variables. 
\end{itemize}

We then add up these contributions, with a weight $\lambda_{\Gamma}$ 
and a power of $N$ given by (with $e(\gamma)$ the number of edges of $\Gamma$, obeying $2e(\Gamma)=qv(\gamma)$ )
\begin{align}
N^{q-k(\Gamma)}\times (N^{-(q-1)})^{v(\Gamma)} \times N^{e(\Gamma)}=N \times N^{-(v(\Gamma)-2)(q/2-1)+1-k(\Gamma)}.
\end{align}

At leading order in $N$, only the Gaussian terms survives ({\it i. e.} the graph $\Gamma$ with $(v(\Gamma)=2$ and $ k(\Gamma)=1))$, except for the matrix model case ($q=2$). In this case, all terms corresponding to connected graphs survive.
Let us emphasize that the variance of the Gaussian distribution of coupling  is thus modified, as a consequence of the non-Gaussian averaging of our model. Remarkably, for $q>2$, this is the only modification at leading order in $N$.

Moreover, the actual value of the covariance (which we denote by $\sigma'$) induced by non Gaussian disorder is most easily computed using a Schwinger-Dyson equation, see \cite{Bonzom:2012cu}. In our context, the latter arises from
\begin{align}
\sum_{i_{1}\dots i_{q}}
\int dJ d \bar{J}\, \frac{\partial}{\partial  \bar{J}_{i_{1}\dots i_{q}}} 
\bigg\{
{J}_{i_{1}\dots i_{q}}
\exp \Bigg[-\Big[\frac{N^{q-1}}{\sigma^{2}}J \bar{J}+V_{N}(J, \bar{J})\Big]
\Bigg]\bigg\}=0.
\end{align}
At large $N$, it leads to the algebraic equation
\begin{align}
1=\frac{\sigma'^{2}}{\sigma^{2}}+
    \sum_{\text{melonic graph $\Gamma$}}\frac{\lambda_{\Gamma} }
{\text{Sym($\Gamma$)}}\,(\sigma')^{v(\Gamma)}
\end{align}

\medskip

Finally, it is interesting to note that this effective action, despite being non local, is invariant under reparametrization (in the IR) at all orders in $1/N$:
\begin{align}
G(t,t')\rightarrow \bigg(\frac{d\phi}{dt}(t)\bigg)^{\Delta}\bigg(\frac{d\phi}{dt'}(t')\bigg)^{\Delta}G(\phi(t),\phi(t')).
\end{align}
Indeed, changing the vertex variables as $t_{v}\rightarrow \phi(t_{v})$, the jacobians exactly cancel  with the rescaling of $G$ since $\Delta=1/q$ and all vertices are are $q$-valent.

\section{A quartic perturbation computation}


In this section we consider the case $q=4$ with a quartic perturbation of the disorder. We explicitly compute the modification of the variance with respect to the Gaussian averaged model and we write down the effective action.

\subsection{The quartic perturbed model}

\begin{figure}
    \centering
    \includegraphics[scale=0.35]{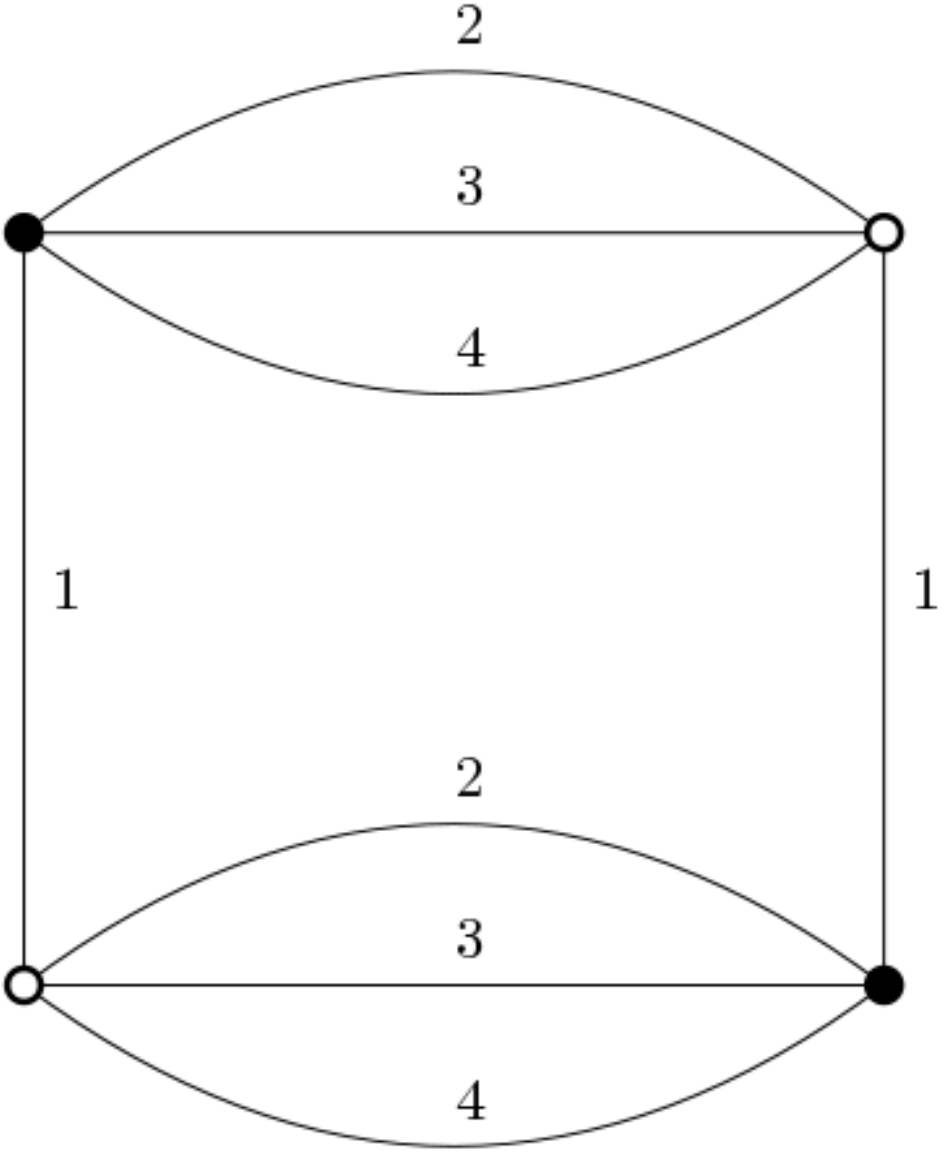}
    \caption{The pillow term for a particular choice of colors of its edges.}
    \label{fig:pillow}
\end{figure}

The action writes:
\begin{equation}
    \mathcal{S}[\psi,\bar{\psi}] = \int \mathrm{d}t \bigg( \sum \limits_{f=1}^{4}\sum \limits_{i=1}
\bar{\psi}_i^{f}\frac{\mathrm{d}}{\mathrm{d}t}\psi_i^{f} - \sum \limits_{i,j,k,l} \bar{J}_{ijkl} \psi_i^{1}\psi_j^{2}\psi_k^{3}\psi_l^{4} -\sum \limits_{i,j,k,l} J_{ijkl} \bar{\psi}_i^{1}\bar{\psi}_j^{2}\bar{\psi}_k^{3}\bar{\psi}_l^{4} \bigg).
\end{equation}
The coupling constant is a random tensor of rank $4$ with the non-Gaussian potential given by: 
\begin{equation}
    V_N(J,\bar{J}) = N^3 \lambda \sum_{c=1}^4 \sum \limits_ {I,K} J_{I}\bar{J}_{I_{\hat{c}}k_c}J_{K}\bar{J}_{K_{\hat{c}}i_c},
\end{equation}
where $I = (i_1,i_2,i_3,i_4)$, $I_{\hat{c}}k_c$ means that $i_c$ is replaced by $k_c$. 
In the tensor model literature, this quartic term is called melonic quartic term or the pillow term, see for example, \cite{Sanchez}, \cite{Pascalie}, \cite{Carrozza} or \cite{Klebanov} or the TASI lectures on large $N$ tensor models \cite{Klebanov2}), see Fig. \ref{fig:pillow}.

We need to integrate over the disorder the replicated generating functional
\begin{align}\label{replicZ}
   \left< \mathrm{Z}^n \right>_J  &= \int\mathcal{D}\psi_i^{f,a}\mathcal{D}\bar{\psi}_i^{f,a}\mathcal{D}J\mathcal{D}\bar{J} \exp \Bigg(- \int \mathrm{d}t \bigg(\sum \limits_{a=1}^{n} \sum \limits_{f=1}^{4}\sum \limits_{i}
\bar{\psi}_{i}^{f,a}\frac{\mathrm{d}}{\mathrm{d}t}\psi_i^{f,a}\bigg) -\frac{N^3}{\sigma^2}\sum_{ijkl}J_{i,j,k,l}\bar{J}_{ijkl} \nonumber \\
&+ \int \mathrm{d}t \bigg(\sum \limits_{a=1}^{n}\sum \limits_{i,j,k,l} \bar{J}_{ijkl} \psi_i^{1,a}\psi_j^{2,a}\psi_k^{3,a}\psi_l^{4,a} +\sum \limits_{a=1}^{n}\sum \limits_{i,j,k,l} J_{ijkl} \bar{\psi}_i^{1,a}\bar{\psi}_j^{2,a}\bar{\psi}_k^{3,a}\bar{\psi}_l^{4,a} \bigg) -V_N(J,\bar{J}) \Bigg).
\end{align}

\subsection{Hubbard-Stratonovich transformation for the disorder}

We start by rewriting the quartic term in $J$ and $\bar J$ using a Hubbard-Stratonovich transformation (or intermediate field representation, see for example \cite{Dartois} or
\cite{Nguyen:2014mga})
\begin{equation}
   e^{-N^3\lambda \sum \limits_ {I,K} J_{I}\bar{J}_{I_{\hat{c}}k_c}J_{K}\bar{J}_{K_{\hat{c}}i_c}} = \sqrt{\frac{N^3}{2\pi}} \int \mathrm{d}M^{(c)} e^{-\frac{N^3}{2}\text{Tr} ((M^{(c)})^2 )- iN^3\big(\frac{\lambda}{2}\big)^{\frac{1}{2}}\sum \limits_{I,j} \bar{J}_{I_{\hat{c}}j}M^{(c)}_{ij}J_I },
\end{equation}
where $M^{(c)}$ is an $N\times N$ Hermitian matrix, for  $c \in \{1,2,3,4\}$. 
In \eqref{replicZ}, keeping only the terms in $J$ and $\bar{J}$, we get
\begin{align}
  \int \mathcal{D}J\mathcal{D}\bar{J}\exp \Bigg(&-\frac{N^{3}}{\sigma^2}\sum \limits_{i,j,k,l}J_{ijkl}\bar{J}_{ijkl}-iN^3\big(\frac{\lambda}{2}\big)^{\frac{1}{2}}\sum_{c=1}^4\sum \limits_{I,j} J_IM^{(c)}_{ij}\bar{J}_{I_{\hat{c}}j}  \nonumber \\
  &+\sum\limits_{a=1}^{n}\sum \limits_{i,j,k,l} \Big( \bar{J}_{ijkl} \psi_i^{1,a}\psi_j^{2,a}\psi_k^{3,a}\psi_l^{4,a} + J_{ijkl} \bar{\psi}_i^{1,a}\bar{\psi}_j^{2,a}\bar{\psi}_k^{3,a}\bar{\psi}_l^{4,a} \Big)\Bigg). \label{Jinterm}
\end{align}
Following \cite{Nguyen:2014mga}, we introduce the notation $\mathcal{M}_c = \mathbb{1}^{\otimes(c-1)} \otimes M^{(c)} \otimes \mathbb{1}^{\otimes(4-c)}$ for $c \in \{1,2,3,4\}$. We can thus rewrite eq. \eqref{Jinterm} as
\begin{equation}
\label{intermediar}
     \int \mathcal{D}J\mathcal{D}\bar{J}\exp \Bigg(-N^{3} \bar{J} \Big( \frac{1}{\sigma^2}\mathbb{1}^{\otimes 4} + i \big(\frac{\lambda}{2}\big)^{\frac{1}{2}}\sum_{c=1}^4\mathcal{M}_c\Big)J +\sum\limits_{a=1}^{n}\sum \limits_{i,j,k,l} \Big( \bar{J}_{ijkl} \psi_i^{1,a}\psi_j^{2,a}\psi_k^{3,a}\psi_l^{4,a} + J_{ijkl} \bar{\psi}_i^{1,a}\bar{\psi}_j^{2,a}\bar{\psi}_k^{3,a}\bar{\psi}_l^{4,a} \Big)\Bigg).
\end{equation}
Then, after rescaling $(J,\bar{J}) \rightarrow N^{\frac{3}{2}}(J,\bar{J})$, we can perform the integral over the disorder. Eq. \eqref{intermediar} thus rewrites:
\begin{equation}
    \frac{(2\pi)^N}{\det\big(\frac{1}{\sigma^2}\mathbb{1}^{\otimes 4} + i \big(\frac{\lambda}{2}\big)^{\frac{1}{2}}\sum_{c=1}^4\mathcal{M}_c\big)}\exp\Bigg(N^{-3}\int \mathrm{d}t_1 \mathrm{d}t_2\sum_{a,b=1}^n\prod_{f=1}^4\psi^{f,a}(t_1) (\frac{1}{\sigma^2}\mathbb{1}^{\otimes 4} + i \big(\frac{\lambda}{2}\big)^{\frac{1}{2}}\sum_{c=1}^4\mathcal{M}_c)^{-1}\prod_{f=1}^4\bar{\psi}^{f,b}(t_2)\Bigg).
\end{equation}
Hence, the replicated generating functional 
\eqref{replicZ}
writes
\begin{align}
\label{replic2}
&\left< \mathrm{Z}^n \right>_J  = \int\mathcal{D}\psi_i^{f,a}\mathcal{D}\bar{\psi}_i^{f,a}\mathcal{D}M^{(c)} \exp \Bigg(- \int \mathrm{d}t \sum \limits_{a=1}^{n} \sum \limits_{f=1}^{4}\sum \limits_{i}
\bar{\psi}_{i}^{f,a}\frac{\mathrm{d}}{\mathrm{d}t}\psi_i^{f,a}  - \text{Tr}\log\Big(\frac{1}{\sigma^2}\mathbb{1}^{\otimes 4} + i \big(\frac{\lambda}{2}\big)^{\frac{1}{2}}\sum_{c=1}^4\mathcal{M}_c\Big) \nonumber \\
&-\frac{1}{2}\sum_{c=1}^4\text{Tr}(\mathcal{M}_c^2) + \frac{1}{N^3}\int \mathrm{d}t_1 \mathrm{d}t_2\sum_{a,b=1}^n\psi^{1,a}\psi^{2,a}\psi^{3,a}\psi^{4,a} (\frac{1}{\sigma^2}\mathbb{1}^{\otimes 4} + i \big(\frac{\lambda}{2}\big)^{\frac{1}{2}}\sum_{c=1}^4\mathcal{M}_c)^{-1}\bar{\psi}^{1,b}\bar{\psi}^{2,b}\bar{\psi}^{3,b}\bar{\psi}^{4,b} \Bigg).
\end{align}
One has:
\begin{align}
    (\frac{1}{\sigma^2}\mathbb{1}^{\otimes 4} + i \big(\frac{\lambda}{2}\big)^{\frac{1}{2}}\sum_{c=1}^4\mathcal{M}_c)^{-1} &= \sigma^2\mathbb{1}^{\otimes 4} +  \sigma^2\sum \limits_{k \geq 1} \frac{(-i\big(\frac{\lambda}{2}\big)^{\frac{1}{2}}\sigma^2)^k}{k!} \big(\sum_{c=1}^4\mathcal{M}_c\big)^k \nonumber \\
    &=\sigma^2\mathbb{1}^{\otimes 4} +  \sigma^2 \sum \limits_{k \geq 1} \frac{(-i\big(\frac{\lambda}{2}\big)^{\frac{1}{2}}\sigma^2)^k}{k!}\sum \limits_{k_1+k_2+k_3+k_4=k}   \binom{k}{k_1,k_2,k_3,k_4} \prod \limits_{c=1}^4\mathcal{M}_c^{k_c}, \label{inversdev}\\
    \log\Big(\frac{1}{\sigma^2}\mathbb{1}^{\otimes 4} +  i \big(\frac{\lambda}{2}\big)^{\frac{1}{2}}\sum_{c=1}^4\mathcal{M}_c\Big) &= -2\log\Big(\sigma\mathbb{1}^{\otimes 4}\Big) +  \log\Big(\mathbb{1}^{\otimes 4} +  i\sigma^2 \big(\frac{\lambda}{2}\big)^{\frac{1}{2}}\sum_{c=1}^4\mathcal{M}_c\Big) \nonumber \\
    &= -2\log\Big(\sigma\mathbb{1}^{\otimes 4}\Big) - \sum \limits_{k \geq 1} \frac{(-i\big(\frac{\lambda}{2}\big)^{\frac{1}{2}}\sigma^2)^k}{k} \sum \limits_{k_1+k_2+k_3+k_4=k}  \binom{k}{k_1,k_2,k_3,k_4} \prod \limits_{c=1}^4\mathcal{M}_c^{k_c} \label{logdev}
\end{align}
Inserting these series in \eqref{replic2} and keeping only the terms in $\mathcal{M}_c$ we have:
\begin{align}
    \label{inter2}
    &-\frac{1}{2}\sum_{c=1}^4\text{Tr}(\mathcal{M}_c^2) +\sum \limits_{k \geq 1} \frac{(-i\big(\frac{\lambda}{2}\big)^{\frac{1}{2}}\sigma^2)^k}{k} \sum \limits_{k_1+k_2+k_3+k_4=k}  \binom{k}{k_1,k_2,k_3,k_4} \text{Tr}\Big(\prod \limits_{c=1}^4\mathcal{M}_c^{k_c}\Big) \nonumber \\
    &+  \frac{\sigma^2}{N^3}\int \mathrm{d}t_1 \mathrm{d}t_2\sum_{a,b=1}^n\psi^{1,a}\psi^{2,a}\psi^{3,a}\psi^{4,a} \sum \limits_{k \geq 1}\frac{(-i\big(\frac{\lambda}{2}\big)^{\frac{1}{2}}\sigma^2)^k}{k!}\sum \limits_{k_1+k_2+k_3+k_4=k}   \binom{k}{k_1,k_2,k_3,k_4}  \prod \limits_{c=1}^4\mathcal{M}_c^{k_c}\bar{\psi}^{1,b}\bar{\psi}^{2,b}\bar{\psi}^{3,b}\bar{\psi}^{4,b}.
\end{align}
Eq. \eqref{inter2}
can be rewritten in the form
\begin{align}
    &-\frac{1}{2}\sum_{c=1}^4\text{Tr}(\mathcal{M}_c^2) +\text{Tr}\Bigg( \sum \limits_{k \geq 1} \sum \limits_{k_1+k_2+k_3+k_4=k}\frac{(-i\big(\frac{\lambda}{2}\big)^{\frac{1}{2}}\sigma^2)^k}{k_1!k_2!k_3!k_4!}\Big((k-1)!\mathbb{1}^{\otimes 4}+\frac{\sigma^2}{N^3} \int \mathrm{d}\mathbf{t}\sum_{a,b=1}^n(\bar{\psi}_{b}^4\cdot\psi_{a}^4)\Big)\prod \limits_{c=1}^4\mathcal{M}_c^{k_c}\Bigg),
    \label{interm-field-action}
\end{align}
where $(\bar{\psi}_{b}^4\cdot\psi_{a}^4) = \bigotimes \limits_{f=1}^4 (\bar{\psi}^{f,b}\cdot\psi^{f,a}) $ with $(\bar{\psi}^{f,b}\cdot\psi^{f,a})_{ij}=\bar{\psi}^{f,b}_i\psi^{f,a}_j$ and $\mathrm{d}\mathbf{t} = \mathrm{d}t_1 \mathrm{d}t_2$. 

We can note that the term proportional to the identity in \eqref{inversdev} contributes to the effective action as the Gaussian part of the disorder. 

\subsection{First order correction of the effective action}

The integration on the intermediate fields $M^{(c)}$ cannot be explictly performed. We can however truncate the series in $\lambda$ and compute perturbatively the first order of the effective action. Keeping only the linear and quadratic terms in the intermediate fields in equation \eqref{interm-field-action}, we get
\begin{equation}
    -\frac{1}{2}\sum_{c=1}^4\text{Tr}(\mathcal{M}_c^2) - i\sigma^2\big(\frac{\lambda}{2}\big)^{\frac{1}{2}} \sum_{c=1}^4\text{Tr}(\mathcal{M}_c) + \frac{\sigma^2}{N^3}\int \mathrm{d}t_1 \mathrm{d}t_2\sum_{a,b=1}^n\psi_{a}^4\Big(-i\big(\frac{\lambda}{2}\big)^{\frac{1}{2}}\sigma^2\sum_{c=1}^4\mathcal{M}_c \Big)\bar{\psi}_{b}^4,
    \label{perturbinterm1}
\end{equation}
where  we have introduced the notations $\psi_{a}^4 =\psi^{1,a}\psi^{2,a}\psi^{3,a}\psi^{4,a}$ and $\bar{\psi}_{b}^4= \bar{\psi}^{1,b}\bar{\psi}^{2,b}\bar{\psi}^{3,b}\bar{\psi}^{4,b}$.
By performing partial traces on the identity part of $\mathcal{M}_c$, equation \eqref{perturbinterm1} can be further simplified:
\begin{equation}
    \int \prod\limits_{c=1}^4\mathrm{d}M^{(c)}\exp\Bigg\{- \frac{N^3}{2}\sum_{c=1}^4\text{Tr}((M^{(c)})^2) - \sum_{c=1}^4\text{Tr}\bigg[\bigg( i\big(\frac{\lambda}{2}\big)^{\frac{1}{2}}\Big(N^3\sigma^2\mathbb{1}+\frac{\sigma^4}{N^3}\int \mathrm{d}\mathbf{t} \sum_{a,b=1}^n(\bar{\psi}_{b}^4\cdot\psi_{a}^4)_c\Big) M^{(c)}\bigg]\Bigg\},
\end{equation}
where $(\bar{\psi}_{b}^4\cdot\psi_{a}^4)_c = \prod \limits_{d \neq c}\text{Tr}(\bar{\psi}^{d,b}\cdot\psi^{d,a}) (\bar{\psi}^{c,b}\cdot\psi^{c,a}) $.
Let us now perform the Gaussian integrals on the intermediate fields. We get:
\begin{align}
    &\exp\Bigg( -\frac{N^3\lambda}{4} \sum\limits_{c=1}^4\text{Tr} \Big(\sigma^2\mathbb{1}+\frac{\sigma^4}{N^6}\int \mathrm{d}\mathbf{t} \sum_{a,b=1}^n(\bar{\psi}_{b}^4\cdot\psi_{a}^4)_c\Big)^2   \Bigg) \nonumber \\
    &= \exp\Bigg( -N^4\lambda\sigma^4 - \frac{2\lambda\sigma^6}{N^3}\int \mathrm{d}\mathbf{t} \sum_{a,b=1}^n\text{Tr}(\bar{\psi}_{b}^4\cdot\psi_{a}^4) - \frac{\lambda\sigma^8}{4N^9} \sum\limits_{c=1}^4\int \mathrm{d}\mathbf{t}\mathrm{d}\mathbf{t}' \sum_{a,b,p,q=1}^n\text{Tr}\big[ (\bar{\psi}_{b}^4\cdot\psi_{a}^4)_c(\bar{\psi}_{p}^4\cdot\psi_{q}^4)_c)\big] \Bigg).
    \label{nongaussEA}
\end{align}
This leads to the the following expression for the effective action
\begin{align}
    \mathcal{S}_{eff}[\psi,\bar{\psi}] &=  \int \mathrm{d}t \sum \limits_{a=1}^{n} \sum \limits_{f=1}^{4}\sum \limits_{i}
    \bar{\psi}_{i}^{f,a}\frac{\mathrm{d}}{\mathrm{d}t}\psi_i^{f,a} -\frac{\sigma^2-2\lambda\sigma^6}{N^3} \int \mathrm{d}\mathbf{t} \sum_{a,b=1}^n\text{Tr}(\bar{\psi}_{b}^4\cdot\psi_{a}^4) \nonumber \\
    &+\frac{\lambda\sigma^8}{4N^9} \sum\limits_{c=1}^4\int \mathrm{d}\mathbf{t}\mathrm{d}\mathbf{t}' \sum_{a,b,p,q=1}^n\text{Tr}\big[ (\bar{\psi}_{b}^4\cdot\psi_{a}^4)_c(\bar{\psi}_{p}^4\cdot\psi_{q}^4)_c)\big], \label{effecactionlambda}
\end{align}
with $\mathrm{d}\mathbf{t}'=\mathrm{d}t_3\mathrm{d}t_4$.
We can directly see the effect of the non-Gaussian perturbation, on the effective action. Taking $\lambda=0$, we recover the action of the model without quartic perturbation in the disorder.
We then define the bi-local fields
\begin{equation}
     {\mathrm{G}}^{ab}_{f}(t_1,t_2)=\frac{1}{N} \sum \limits_{i}
     \psi^{f,b}_{i}(t_1)
     \bar{\psi}^{f,a}_{i}(t_2),
\end{equation}
and introduce the Lagrange multipliers ${\Sigma}^{ab}_{f}(t_1,t_2)$ :
\begin{equation}
\int \mathcal{D}{\Sigma}^{ab}_f \exp{\Bigg(-N\int \mathrm{d}\mathbf{t}\sum \limits_{a,b=1}^{n} \sum \limits_{f=1}^{4}{\Sigma}^{ab}_f(\mathbf{t})\bigg( {\mathrm{G}}^{ab}_f(\mathbf{t}) - \frac{1}{N}  \sum \limits_{i} 
\psi^{f,b}_{i}(t_1)
     \bar{\psi}^{f,a}_{i}(t_2)
\bigg) \Bigg)}.
\end{equation}
The effective action \eqref{effecactionlambda} rewrites as:
\begin{align}
    &\mathcal{S}_{eff}[\psi,\bar{\psi},{\mathrm{G}},{\Sigma}] = \int \mathrm{d}\mathbf{t} \sum \limits_{a,b=1}^{n} \sum \limits_{f=1}^{4}\sum \limits_{i}
    \bar{\psi}_{i}^{f,a}\big(\delta_{ab}\delta(t_1-t_2)\partial_{t}-{\Sigma}^{ab}_f(\mathbf{t}) \big)\psi_i^{f,a} -N(\sigma^2-2\lambda\sigma^6) \int \mathrm{d}\mathbf{t} \sum_{a,b=1}^n\prod\limits_{f=1}^4{\mathrm{G}}^{ab}_{f}(\mathbf{t}) \nonumber \\
    &+N \sum \limits_{a,b=1}^{n}\int \mathrm{d}\mathbf{t}\sum \limits_{f=1}^{4} {\Sigma}^{ab}_f(\mathbf{t}) {\mathrm{G}}^{ab}_f(\mathbf{t})+\frac{\lambda\sigma^8}{4N} \int \mathrm{d}\mathbf{t}\mathrm{d}\mathbf{t}' \sum_{a,b,p,q=1}^n\sum\limits_{c=1}^4{\mathrm{G}}^{ap}_{c}(t_1,t_4){\mathrm{G}}^{qb}_{c}(t_3,t_2)\prod\limits_{f \neq c}{\mathrm{G}}^{ab}_{f}(\mathbf{t}){\mathrm{G}}^{qp}_{f}(\mathbf{t}').
\end{align}
We now perform the Gaussian integral on the fermionic fields
\begin{align}
    &\left< \mathrm{Z}^n \right>_J  = \int\mathcal{D}\psi_i^{f,a}\mathcal{D}\bar{\psi}_i^{f,a}\mathcal{D}{\mathrm{G}}^{ab}_{f} \mathcal{D}{\Sigma}^{ab}_f \exp\big(-\mathcal{S}_{eff}[\psi,\bar{\psi},{\mathrm{G}},{\Sigma}]\big)  \\
    &=\int\mathcal{D}{\mathrm{G}}^{ab}_{f} \mathcal{D}{\Sigma}^{ab}_f \exp\Bigg(\frac{N}{2} \sum \limits_{a,b=1}^{n}\sum \limits_{f=1}^{4} \log{\det{\big(\delta_{ab}\delta(t_1-t_2)\partial_{t}-{\Sigma}^{ab}_f(t_1,t_2) \big)}}+N(\sigma^2-2\lambda\sigma^6) \int \mathrm{d}\mathbf{t} \sum_{a,b=1}^n\prod\limits_{f=1}^4{\mathrm{G}}^{ab}_{f}(\mathbf{t}) \nonumber  \\
    &-N \sum \limits_{a,b=1}^{n}\int \mathrm{d}\mathbf{t}\sum \limits_{f=1}^{4} {\Sigma}^{ab}_f(\mathbf{t}) {\mathrm{G}}^{ab}_f(\mathbf{t})-\frac{\lambda\sigma^8}{4N} \int \mathrm{d}\mathbf{t}\mathrm{d}\mathbf{t}' \sum_{a,b,p,q=1}^n\sum\limits_{c=1}^4{\mathrm{G}}^{ap}_{c}(t_1,t_4){\mathrm{G}}^{qb}_{c}(t_3,t_2)\prod\limits_{f \neq c}{\mathrm{G}}^{ab}_{f}(\mathbf{t}){\mathrm{G}}^{qp}_{f}(\mathbf{t}')\Bigg). \nonumber 
\end{align}
We then assume a symmetric saddle point for the replicas and get the effective action:
\begin{align}
\label{effectaction}
    &\frac{\mathcal{S}_{eff}[{\mathrm{G}},{\Sigma}]}{N} = - \sum \limits_{f=1}^{4} \log{\det{\big(\delta(t_1-t_2)\partial_{t}-{\Sigma}_f(t_1,t_2) \big)}}-(\sigma^2-2\lambda\sigma^6) \int \mathrm{d}\mathbf{t} \prod\limits_{f=1}^4{\mathrm{G}}_{f}(\mathbf{t}) \nonumber  \\
    &+\int \mathrm{d}\mathbf{t} \sum \limits_{f=1}^{4}{\Sigma}_f(\mathbf{t}) {\mathrm{G}}_f(\mathbf{t})-\frac{\lambda\sigma^8}{4N^2} \int \mathrm{d}\mathbf{t}\mathrm{d}\mathbf{t}' \sum\limits_{c=1}^4{\mathrm{G}}_{c}(t_1,t_4){\mathrm{G}}_{c}(t_3,t_2)\prod\limits_{f \neq c}{\mathrm{G}}_{f}(\mathbf{t}){\mathrm{G}}_{f}(\mathbf{t}').
\end{align}
Let us emphasize that the effective action formula above implies that 
the variance of the Gaussian disorder is modified by the non-Gaussian perturbation:
\begin{equation}
\label{rezultat}
\sigma^2 \rightarrow \sigma^2-2\lambda\sigma^6.    
\end{equation}
Moreover, let us notice that 
every term in \eqref{effectaction}
is of order $1$, except for the last term  which is of order $\frac{1}{N^2}$, as expected from the universality result. This term can be represented graphically as in Fig \ref{fig:effpillow}.

\section{Gross-Rosenhaus SYK model with non-Gaussian disorder}

In this section we follow the steps of the calculation of the previous section and we compute the modification of the covariance and of the effective action for a quartic melonic perturbation of the disorder of the Gross-Rosenhaus SYK model. 

As already mentioned in the Introduction, the Gross-Rosenhaus model
\cite{Gross}
is a generalisation of the 
SYK model 
containing $f$ flavors of  fermions,  with $N_a$ fermions of flavor $a$, appearing $q_a$ times in the interaction, so that $q = \sum \limits_{a=1}^f q_a$. The complex model treated in the previous section can thus be seen as a particular case of a complex version of the  Gross-Rosenhaus model treated in this section.

The action of the model writes 
\begin{equation}
    \mathcal{S} = \int \mathrm{d}t \Bigg(\sum \limits_{a=1}^{f}\sum \limits_{i=1}^{N_a}
\psi_i^{a}\frac{\mathrm{d}}{\mathrm{d}t}\psi_i^{a} + \frac{(i)^{\frac{q}{2}}}{\prod_{a=1}^f q_a!}\sum \limits_{I} J_{I} \prod_{a=1}^f \prod _{p=1}^{q_a}(\psi_{i^a_{p}}^{a}) \Bigg),
\end{equation}
where $I = i^1_1, \hdots, i^1_{q_1}, \hdots, i^f_{1}, \hdots, i^f_{q_f} $. The coupling tensor $J$ is now antisymmetric under permutations of indices in the same family of flavors, with the probability distribution
\begin{equation}
    P(J) =C \exp{\Bigg(-\frac{1}{2\sigma^2N}\prod_a\frac{ N_a^{q_a}}{(q_a-1)!}\sum \limits_{I}J_{I}^2 - \frac{\lambda}{4} \frac{\prod_a N_a^{q_a}}{N} \sum_{c=1}^q \sum \limits_ {I,K} J_{I}J_{I_{\hat{c}}k_c}J_{K}J_{K_{\hat{c}}i_c}\Bigg)},
\end{equation}
where $N= \sum \limits_a N_a$. We use again the intermediate field
\begin{equation}
   e^{- \frac{\lambda}{4} \frac{\prod_a N_a^{q_a}}{N} \sum \limits_ {I,K}  J_{I}J_{I_{\hat{c}}k_c}J_{K}J_{K_{\hat{c}}i_c}} \propto \int \mathrm{d}M^{(c)} e^{-\frac{1}{2}\frac{\prod_a N_a^{q_a}}{N}\text{Tr} ((M^{(c)})^2 )- i\frac{\prod_a N_a^{q_a}}{N}\big(\frac{\lambda}{2}\big)^{\frac{1}{2}}\sum \limits_{I,j} J_{I_{\hat{c}}j}M^{(c)}_{ij}J_I },
\end{equation}
where $M^{(c)}$ is a symmetric real $N_c \times N_c$ matrix. Using the replica trick and keeping the terms in $J$ we have
\begin{align}
  \int \mathcal{D}J  \exp \Bigg( &-\prod_a\frac{ N_a^{q_a}}{(q_a-1)!}\sum \limits_{I}\frac{J_{I}^2}{2\sigma^2N} -i\frac{\prod_a N_a^{q_a}}{N}\big(\frac{\lambda}{2}\big)^{\frac{1}{2}}\sum \limits_{c=1}^q\sum \limits_{I,j} J_{I_{\hat{c}}j}M^{(c)}_{ij}J_I \nonumber \\
  & - \frac{(i)^{\frac{q}{2}}}{\prod_{a=1}^f q_a!} \int \mathrm{d}t \sum \limits_{r=1}^n\sum \limits_{I} J_{I} \prod_{a=1}^f \prod _{p=1}^{q_a}(\psi_{i^{a}_{p}}^{a,r}) \Bigg).
\end{align}
As above, let $\mathcal{M}_c = \mathbb{1}^{\otimes(c-1)} \otimes M^{(c)} \otimes \mathbb{1}^{\otimes(q-c)}$, where $\mathbb{1}$ is implicitly the identity $N_a \times N_a$ matrix, for $a=1,\hdots,f$. 
One then has
\begin{align}
  &\int \mathcal{D}J  \exp{\Bigg( -\frac{\prod_a N_a^{q_a}}{2N}J\bigg(\frac{1}{\sigma^2}\prod_{a=1}^f\frac{1}{(q_a-1)!}\mathbb{1}^{\otimes q} +i\big(2\lambda\big)^{\frac{1}{2}}\sum \limits_{c=1}^q \mathcal{M}_c\bigg) J - \frac{(i)^{\frac{q}{2}}}{\prod_{a=1}^f q_a!} \int \mathrm{d}t \sum \limits_{r=1}^n\sum \limits_{I} J_{I} \prod_{a=1}^f \prod_{p=1}^{q_a}(\psi_{i^{a}_{p}}^{a,r}) \Bigg)} \nonumber \\
  &\propto \bigg(\det{\Big(\frac{1}{\sigma^2}\prod_{a=1}^f\frac{1}{(q_a-1)!}\mathbb{1}^{\otimes q} +i\big(2\lambda\big)^{\frac{1}{2}}\sum \limits_{c=1}^q \mathcal{M}_c\Big)}\bigg)^{-\frac{1}{2}} \nonumber \\
  & \exp{\Bigg(\frac{i^q N}{2\prod_{a=1}^f N_a^{q_a} (q_a!)^2} \int \mathrm{d}t_1 \mathrm{d}t_2 \sum \limits_{r,s=1}^n \prod_{a=1}^f \prod_{p=1}^{q_a}(\psi_{p}^{a,r}) \left(\frac{1}{\sigma^2}\prod_{a=1}^f\frac{1}{(q_a-1)!}\mathbb{1}^{\otimes q} +i\big(2\lambda\big)^{\frac{1}{2}}\sum \limits_{c=1}^q \mathcal{M}_c\right)^{-1} (\psi_{p}^{a,s}) \Bigg)}.
\end{align}
The replicated generating functional thus writes:
\begin{align}
\left< \mathrm{Z}^n \right>_J  
    & = \int\mathcal{D}\psi_i^{f,a}\mathcal{D}M^{(c)} \exp \left[ -\int \mathrm{d}t \sum_{r=1}^n \sum \limits_{a=1}^{f}\sum \limits_{i=1}^{N_a} \psi_i^{a,r}\frac{\mathrm{d}}{\mathrm{d}t}\psi_i^{a,r} \right. \nonumber \\
    & -\frac{1}{2} \text{Tr} \log\Big(\frac{1}{\sigma^2}\prod_{a=1}^f\frac{1}{(q_a-1)!}\mathbb{1}^{\otimes q} +i\big(2\lambda\big)^{\frac{1}{2}}\sum \limits_{c=1}^q \mathcal{M}_c\Big) -\frac{1}{2}\frac{\prod_a N_a^{q_a}}{N} \sum_{c=1}^q \text{Tr} ((M^{(c)})^2 ) \nonumber \\
    & \left.  +  \frac{i^q N}{2\prod_{a=1}^f N_a^{q_a} (q_a!)^2} \int \mathrm{d}t_1 \mathrm{d}t_2  \sum \limits_{r,s=1}^n \prod_{a=1}^f \prod_{p=1}^{q_a}(\psi_{p}^{a,r}) \left(\frac{1}{\sigma^2}\prod_{a=1}^f\frac{1}{(q_a-1)!}\mathbb{1}^{\otimes q} +i\big(2\lambda\big)^{\frac{1}{2}}\sum \limits_{c=1}^q \mathcal{M}_c\right)^{-1} (\psi_{p}^{a,s})   \right].
\end{align}
As above, we now write the following series:
\begin{align}
    & \left(\frac{1}{\sigma^2}\prod_{a=1}^f\frac{1}{(q_a-1)!}\mathbb{1}^{\otimes q} +i\big(2\lambda\big)^{\frac{1}{2}}\sum \limits_{c=1}^q \mathcal{M}_c\right)^{-1} \nonumber \\
    & = \sigma^2\prod_{a=1}^f(q_a-1)!\mathbb{1}^{\otimes q} 
    + \sigma^2\prod_{a=1}^f(q_a-1)!\sum_{k \ge  1} \left( -i\sigma^2 \big(2\lambda\big)^{\frac{1}{2}} \prod_{a=1}^f(q_a-1)! \right)^k \sum_{\sum_i^q k_i=k}\binom{k}{k_1, \dots ,k_q}\prod_{c=1}^q \mathcal{M}_c^{k_c},  \\
    & \log \Big(\frac{1}{\sigma^2}\prod_{a=1}^f\frac{1}{(q_a-1)!}\mathbb{1}^{\otimes q} +i\big(2\lambda\big)^{\frac{1}{2}}\sum \limits_{c=1}^q \mathcal{M}_c\Big) \nonumber \\
    & = -2\log(\sigma\mathbb{1}^{\otimes q})-\sum_{a=1}^f \log\left((q_a-1)!\mathbb{1}^{\otimes q}\right) + \log\left( \mathbb{1}^{\otimes q} +i\big(2\lambda\big)^{\frac{1}{2}} \sigma^2 \prod_{a=1}^f (q_a-1)! \sum \limits_{c=1}^q \mathcal{M}_c \right) \nonumber \\
    & = -2\log(\sigma \mathbb{1}^{\otimes q})-\sum_{a=1}^f \log\left((q_a-1)!\mathbb{1}^{\otimes q}\right) - \sum_{k \ge 1} \frac{\left( -i\big(2\lambda\big)^{\frac{1}{2}} \sigma^2 \prod_{a=1}^f (q_a-1)! \right)^k}{k} \sum_{\sum_i^q k_i=k}\binom{k}{k_1, \dots ,k_q}\prod_{c=1}^q \mathcal{M}_c^{k_c}.
\end{align}
This leads to:
\begin{align}
    & -\frac{1}{2}\frac{\prod_a N_a^{q_a}}{N} \sum_{c=1}^q \text{Tr} ((M^{(c)})^2) + \frac{1}{2}\sum_{k \ge 1} \frac{\left( -i\big(2\lambda\big)^{\frac{1}{2}} \sigma^2 \prod_{a=1}^f (q_a-1)! \right)^k}{k} \sum_{\sum_i^q k_i=k}\binom{k}{k_1, \dots ,k_q} \text{Tr}\left(\prod_{c=1}^q \mathcal{M}_c^{k_c}\right) \nonumber \\
    & +  \frac{i^q N \sigma^2}{2\prod_{a=1}^f N_a^{q_a} q_a (q_a!)}  \sum_{k \ge  1} \left( -i\sigma^2 \big(2\lambda\big)^{\frac{1}{2}} \prod_{a=1}^f(q_a-1)! \right)^k \sum_{\sum_i^q k_i=k}\binom{k}{k_1, \dots ,k_q} \int \mathrm{d}\mathbf{t} \text{Tr}\left((\psi_r^q\cdot \psi_s^q)  \prod_{c=1}^q \mathcal{M}_c^{k_c} \right),
\end{align}
where we used the notation $(\psi_r^q \cdot \psi_s^q)= \sum \limits_{r,s=1}^n \bigotimes \limits_{a=1}^f \bigotimes \limits_{p=1}^{q_a}(\psi_{p}^{a,r} \cdot \psi_{p}^{a,s})$ and $(\psi_{p}^{a,r} \cdot \psi_{p}^{a,s})_{ij} = \psi_{i^{a}_{p}}^{a,r} \psi_{j^{a}_{p}}^{a,s}$. For the sake of simplicity, we can rewrite this term in the more compact form
\begin{equation}
\label{Int.Field_Action_Gross-Ros}
    -\frac{1}{2}\frac{\prod_a N_a^{q_a}}{N} \sum_{c=1}^q \text{Tr} ((M^{(c)})^2) + \frac{1}{2}\text{Tr}\left[ \sum_{k \ge 1} \left( -i\big(2\lambda\big)^{\frac{1}{2}} \sigma^2 \prod_{a=1}^f (q_a-1)! \right)^k \sum_{\sum_i^q k_i=k}\binom{k}{k_1, \dots ,k_q} \hat{A}_k \prod_{c=1}^q \mathcal{M}_c^{k_c} \right] ,
\end{equation}
where we denoted 
\begin{equation}
    \hat{A}_k=\frac{\mathbb{1}^{\otimes q}}{k} + \frac{i^q N \sigma^2}{\prod_{a=1}^f N_a^{q_a} q_a (q_a!)}\int \mathrm{d}\mathbf{t} (\psi_r^q\cdot \psi_s^q).
\end{equation}
Keeping only the terms in $\sqrt{\lambda}$ in
\eqref{Int.Field_Action_Gross-Ros}, we get:
\begin{equation}
    -\frac{1}{2}\frac{\prod_a N_a^{q_a}}{N} \sum_{c=1}^q \text{Tr} ((M^{(c)})^2) - i\sigma^2\prod_{a=1}^f (q_a-1)!\sqrt{\frac{\lambda}{2}} \sum_{c=1}^q \text{Tr} \left( \hat{A}_1 \mathcal{M}_c \right).
\end{equation}
This allows to perform the Gaussian integral over the intermediate fields:
\begin{align}
    & \int \prod_{c=1}^q \mathrm{d}M^{(c)} \exp{\left( -\frac{1}{2}\frac{\prod_a N_a^{q_a}}{N} \sum_{c=1}^q \text{Tr} ((M^{(c)})^2) - i\sigma^2\prod_{a=1}^f (q_a-1)!\sqrt{\frac{\lambda}{2}}\sum_{c=1}^q \text{Tr} \left(  \hat{A}_1 \mathcal{M}_c \right)\right)}  \nonumber \\
    & \propto \exp{\left( -\frac{i^q \lambda N^2 \sigma^6}{2\prod_{a=1}^f N_a^{2q_a} q_a^3 (q_a!)^{-1}}\int \mathrm{d}\mathbf{t} \text{Tr}(\psi_r^q\cdot \psi_s^q) - 
    \frac{(-1)^q \lambda N^3 \sigma^8}{4\prod_{a=1}^f N_a^{3q_a} q_a^4}\sum_{c=1}^f\sum_{k=1}^{q_c}\int \mathrm{d}\mathbf{t}\mathrm{d}\mathbf{t'} \text{Tr}\big[(\psi_r^q\cdot \psi_s^q)_{c,k}(\psi_u^q\cdot \psi_v^q)_{c,k} \big]\right)},
\end{align}
where $(\psi_r^q \cdot \psi_s^q)_{c,k}= \sum \limits_{r,s=1}^n \prod \limits_{a\neq c} \prod\limits_{p\neq k}\text{Tr}\big[(\psi_{p}^{a,r} \cdot \psi_{p}^{a,s})\big](\psi_{k}^{c,r} \cdot \psi_{k}^{c,s})$. 
This leads to 
the following expression for the effective action:
\begin{align}
\label{eff.action_Gross-Ros}
    \mathcal{S}_{eff} [\psi]
    & = \int \mathrm{d}t \sum \limits_{a=1}^{f} \sum_{r=1}^n \sum \limits_{i=1}^{N_a} \psi_i^{a,r}\frac{\mathrm{d}}{\mathrm{d}t}\psi_i^{a,r} -
    \frac{i^q N}{2 \prod_{a=1}^f N_a^{q_a} q_a q_a!}\left( \sigma^2 - \lambda N \sigma^6 \prod_{a=1}^f \frac{(q_a!)^2}{q_a^2 N_a^{q_a}} \right) \int \mathrm{d}\mathbf{t} \text{Tr}(\psi_r^q\cdot \psi_s^q)  \nonumber \\
    & + \frac{(-1)^q \lambda N^3 \sigma^8}{4\prod_{a=1}^f N_a^{3q_a} q_a^4}\sum_{c=1}^f\sum_{k=1}^{q_c}\int \mathrm{d}\mathbf{t}\mathrm{d}\mathbf{t'} \text{Tr}\big[(\psi_r^q\cdot \psi_s^q)_{c,k}(\psi_u^q\cdot \psi_v^q)_{c,k}\big].
\end{align}
In order to evaluate the fermionic integral in the expression of the replicated generating function, we introduce the bi-local fields
\begin{equation}
    {G}_a^{rs}(t_1,t_2)=\frac{1}{N_a}\sum_{i=1}^{N_a} \psi_i^{a,r}(t_1)\psi_i^{a,s}(t_2)
\end{equation}
and the Lagrange multipliers ${\Sigma}_a^{rs}(t_1,t_2)$:
\begin{equation}
    \int \mathcal{D}{\Sigma}_a^{rs}\exp{\left(- \int \mathrm{d}\mathbf{t}\sum_{r,s=1}^n\sum_{a=1}^f \frac{N_a}{2} {\Sigma}_a^{rs}(t_1,t_2) \left( {G}_a^{rs}(t_1,t_2) - \frac{1}{N_a}\sum_{i=1}^{N_a} \psi_i^{a,r}\psi_i^{a,s} \right) \right)}.
\end{equation}
The effective action \eqref{eff.action_Gross-Ros} then writes:
\begin{align}
    &\mathcal{S}_{eff} [\psi,{G},{\Sigma}]
    = \int \mathrm{d}\mathbf{t} \sum \limits_{a=1}^{f} \sum_{r,s=1}^n \sum \limits_{i=1}^{N_a} \psi_i^{a,r}\left( \delta_{rs}\delta(t_1-t_2)\partial_t - {\Sigma}_a^{rs}(\mathbf{t})\right)\psi_i^{a,s} + \int \mathrm{d}\mathbf{t}\sum_{r,s=1}^n\sum_{a=1}^f \frac{N_a}{2} {\Sigma}_a^{rs}(\mathbf{t}) {G}_a^{rs}(\mathbf{t}) \nonumber \\
    & - \frac{i^q N}{2 \prod_{a=1}^f  q_a q_a!}\left( \sigma^2 - \lambda N \sigma^6 \prod_{a=1}^f \frac{(q_a!)^2}{q_a^2 N_a^{q_a}} \right) \int\mathrm{d}\mathbf{t} \sum_{r,s=1}^n \prod_{a=1}^f \big({G}_a^{rs}(\mathbf{t})\big)^{q_a} \nonumber \\
    & + \frac{(-1)^q \lambda N^3 \sigma^8}{4\prod_{a=1}^f N_a^{q_a} q_a^4}\int \mathrm{d}\mathbf{t}\mathrm{d}\mathbf{t'} \sum_{r,s,u,v=1}^n \sum_{c=1}^f q_c {G}_c^{su}(t_2,t_3){G}_c^{rv}(t_1,t_4)\big({G}_c^{rs}(\mathbf{t})\big)^{q_c-1}\big({G}_c^{uv}(\mathbf{t'})\big)^{q_c-1}\prod_{a \neq c} \big({G}_a^{rs}(\mathbf{t})\big)^{q_a} \big({G}_a^{uv}(\mathbf{t'})\big)^{q_a}.
\end{align}
This allows to perform the Gaussian integral on the fermionic fields:
\begin{align}
    &\left<\mathrm{Z}^n\right>_J  
    = \int \mathcal{D}\psi \mathcal{D} {G} \mathcal{D} {\Sigma} e^{-\mathcal{S}_{eff} [\psi, {G}, {\Sigma}]} \nonumber \\
    & = \int \mathcal{D}\psi \mathcal{D} {G} \mathcal{D} {\Sigma} 
    \exp \left( -\int \mathrm{d}\mathbf{t} \sum_{a=1}^{f} \sum_{r,s=1}^n \frac{N_a}{2} \log\det\left( \delta_{rs}\delta(t_1-t_2)\partial_t -  {\Sigma}_a^{rs}(\mathbf{t})\right)  \right. \nonumber \\
    & - \int \mathrm{d}\mathbf{t} \sum_{r,s=1}^n\sum_{a=1}^f \frac{N_a}{2}  {\Sigma}_a^{rs}(\mathbf{t})  {G}_a^{rs}(\mathbf{t}) + \frac{i^q N}{2 \prod_{a=1}^f  q_a q_a!}\left( \sigma^2 - \lambda N \sigma^6 \prod_{a=1}^f \frac{(q_a!)^2}{q_a^2 N_a^{q_a}} \right) \int\mathrm{d}\mathbf{t} \sum_{r,s=1}^n \prod_{a=1}^f \big( {G}_a^{rs}(\mathbf{t})\big)^{q_a}  \nonumber \\
    & - \left. \frac{(-1)^q \lambda N^3 \sigma^8}{4\prod_{a=1}^f N_a^{q_a} q_a^4}\int \mathrm{d}\mathbf{t}\mathrm{d}\mathbf{t'} \sum_{r,s,u,v=1}^n \sum_{c=1}^f q_c {G}_c^{su}(t_2,t_3) {G}_c^{rv}(t_1,t_4)\big( {G}_c^{rs}(\mathbf{t})\big)^{q_c-1}\big( {G}_c^{uv}(\mathbf{t'})\big)^{q_c-1}\prod_{a \neq c} \big( {G}_a^{rs}(\mathbf{t})\big)^{q_a} \big( {G}_a^{uv}(\mathbf{t'})\big)^{q_a}\right).
\end{align}
Assuming a symmetric saddle point for the replicas, 
we get the following expression for the effective action:
\begin{align}
    \mathcal{S}_{eff} [ {G}, {\Sigma}]
    & = \int \mathrm{d}\mathbf{t} \sum_{a=1}^{f} \frac{N_a}{2} \log\det\left( \delta(t_1-t_2)\partial_t -  {\Sigma}_a(\mathbf{t})\right) + \int \mathrm{d}\mathbf{t} \sum_{a=1}^f \frac{N_a}{2}  {\Sigma}_a(\mathbf{t})  {G}_a(\mathbf{t})  \nonumber \\
    & - \frac{i^q N}{2 \prod_{a=1}^f  q_a q_a!}\left( \sigma^2 - \lambda N \sigma^6 \prod_{a=1}^f \frac{(q_a!)^2}{q_a^2 N_a^{q_a}} \right) \int\mathrm{d}\mathbf{t} \prod_{a=1}^f \big( {G}_a(\mathbf{t})\big)^{q_a}  \nonumber \\
    & + \frac{(-1)^q \lambda N^3 \sigma^8}{4\prod_{a=1}^f N_a^{q_a} q_a^4}\int \mathrm{d}\mathbf{t}\mathrm{d}\mathbf{t'} \sum_{c=1}^f q_c {G}_c(t_2,t_3) {G}_c(t_1,t_4)\big( {G}_c(\mathbf{t})\big)^{q_c-1}\big( {G}_c(\mathbf{t'})\big)^{q_c-1}\prod_{a \neq c} \big( {G}_a(\mathbf{t})\big)^{q_a} \big( {G}_a(\mathbf{t'})\big)^{q_a}.
\end{align}
Let us emphasize  that the Gaussian variance is now modified by the non-Gaussian perturbation:
\begin{equation}
    \sigma^2 \rightarrow  \sigma^2 - \lambda N \sigma^6 \prod_{a=1}^f \frac{(q_a!)^2}{q_a^2 N_a^{q_a}}.
\end{equation}
This expression is a generalization of the modification \eqref{rezultat} of the Gaussian variance of the complex model treated in the previous sections.

\section{Concluding remarks}

In this paper we have investigated the effects of non-Gaussian average over the random couplings $J$ in a complex SYK model, as well as in a (real) SYK generalization proposed by Gross and Rosenhaus. To our knowledge, this is the first study of the effects of the relaxation of the Gaussianity condition in SYK models when no double scaling limit is taken.

An interesting perspective appears to us to be the investigation of the effects of such a perturbation from Gaussianity in the case of $q=2$ (fermions with  a random mass matrix) and in the case 
of the real SYK model - a first step towards this latter case having been already made in this paper
(since the real SYK model is a particular case of the Gross-Rosenhaus model). 
The main technical complication 
in this latter case 
comes from the fact that one has to deal with graphs which are not necessary bipartite - the removal and reconnection of edges of these graphs (which is the main technical ingredient of our approach) being much more involved.

It would thus be interesting to check weather or not in this case also, non-Gaussian perturbation leads to a modification of the variance of the Gaussian distributions of the couplings $J$ at leading order in $N$, as we proved to be the case for the complex version of the SYK model studied here.

\section*{Acknowledgements}
The authors warmly acknowledge St\'ephane Dartois for various discussions on the Dyson-Schwinger equations. The authors are partially supported by the CNRS Infiniti ModTens grant. A. Tanasa is partially supported by the PN 09 37 01 02 grant.

\appendix

\section{Dyson-Schwinger eq. for the intermediate field}

In this appendix we construct the Dyson-Schwinger equations for the matrix intermediate field used in the paper. As already announced in the Introduction, our calculations follow the lines of \cite{Nguyen:2014mga}. The following subsection deals with the complex SYK model and the last subsection deals with the real Gross-Rosenhaus SYK generalization.

\subsection{Complex SYK}

We first  perform the following change of variables: 
\begin{equation}
M^{(c)} \rightarrow \alpha \mathbb{1} + \frac{1}{N}M^{(c)}.
\end{equation}
The effective action for the intermediate field in equation \eqref{interm-field-action} leads to the following expression
\begin{align}
    & -\frac{N}{2}\sum_{c=1}^4\text{Tr}(M^{(c)2})-\alpha N^2\sum_{c=1}^4\text{Tr}(M^{(c)}) + \nonumber \\
    & +\text{Tr}\Bigg( \sum \limits_{k \geq 1} \sum \limits_{k_1+k_2+k_3+k_4=k}\frac{(-i\big(\frac{\lambda}{2}\big)^{\frac{1}{2}}\sigma^2)^k}{k_1!k_2!k_3!k_4!}\Big((k-1)!\mathbb{1}^{\otimes 4}+\frac{\sigma^2}{N^3} \int \mathrm{d}\mathbf{t}\sum_{a,b=1}^n(\bar{\psi}_{b}^4\cdot\psi_{a}^4)\Big)\prod \limits_{c=1}^4\Big(\alpha \mathbb{1}^{\otimes 4}+\frac{1}{N}\mathcal{M}_c\Big)^{k_c}\Bigg)= \nonumber \\
    & =-\frac{N}{2}\sum_{c=1}^4\text{Tr}(M^{(c)2})-\alpha N^2\sum_{c=1}^4\text{Tr}(M^{(c)}) + \nonumber \\
    & +\text{Tr}\Bigg( \sum \limits_{k \geq 1} \sum \limits_{k_1+k_2+k_3+k_4=k}\frac{(-i\big(\frac{\lambda}{2}\big)^{\frac{1}{2}}\sigma^2\alpha)^k}{k_1!k_2!k_3!k_4!}\Big((k-1)!\mathbb{1}^{\otimes 4}+\frac{\sigma^2}{N^3} \int \mathrm{d}\mathbf{t}\sum_{a,b=1}^n(\bar{\psi}_{b}^4\cdot\psi_{a}^4)\Big)\prod \limits_{c=1}^4 \sum_{p_c=0}^{k_c} \binom{k_c}{p_c}\frac{\mathcal{M}_c^{p_c}}{(\alpha N)^{p_c}}\Bigg),
\end{align}
where we recall the notation $\mathcal{M}_c = \mathbb{1}^{\otimes(c-1)} \otimes M^{(c)} \otimes \mathbb{1}^{\otimes(4-c)}$. 
Using the expression above of the action, 
we can now derive the Dyson-Schwinger equations (recall that these equations can be derived by exploiting the fact that the integration of a total derivative is vanishing):
\begin{align}
    & 0=\sum_{ij}\int\prod_{c=1}^4\mathrm{d}M^{(c)} \frac{\partial}{\partial M_{ij}^{(d)}} \Bigg[\Big( M^{(d)}\Big)_{ij}^q \exp \Bigg(
    -\frac{N}{2}\sum_{c=1}^4\text{Tr}(M_c^2)-\alpha N^2\sum_{c=1}^4\text{Tr}(M_c) + \nonumber \\
    & +\text{Tr}\Bigg( \sum \limits_{k \geq 1} \sum \limits_{k_1+k_2+k_3+k_4=k}\frac{(-i\big(\frac{\lambda}{2}\big)^{\frac{1}{2}}\sigma^2\alpha)^k}{k_1!k_2!k_3!k_4!}\Big((k-1)!\mathbb{1}^{\otimes 4}+\frac{\sigma^2}{N^3} \int \mathrm{d}\mathbf{t}\sum_{a,b=1}^n(\bar{\psi}_{b}^4\cdot\psi_{a}^4)\Big)\prod \limits_{c=1}^4 \sum_{p_c=0}^{k_c} \binom{k_c}{p_c}\frac{\mathcal{M}_c^{p_c}}{(\alpha N)^{p_c}}\Bigg)\Bigg)\Bigg].
    \label{loop-eq}
\end{align}
This leads to:
\begin{align}
    & 0=\left\langle\sum_{i=0}^{q-1}\text{Tr}\left(M^{(d)i}\right)\text{Tr}\left(M^{(d){q-i-1}}\right)\right\rangle-N\left\langle\text{Tr}\left(M^{(d){q+1}}\right)\right\rangle-\alpha N^2\left\langle\text{Tr}\left(M^{(d)q}\right)\right\rangle+ \nonumber \\
    & +\left\langle \text{Tr}\left[ \sum \limits_{k \geq 1} \sum \limits_{k_1+k_2+k_3+k_4=k}\frac{(-i\big(\frac{\lambda}{2}\big)^{\frac{1}{2}}\sigma^2\alpha)^k}{k_1!k_2!k_3!k_4!}\Big((k-1)!\mathbb{1}^{\otimes 4}+\frac{\sigma^2}{N^3} \int \mathrm{d}\mathbf{t}\sum_{a,b=1}^n(\bar{\psi}_{b}^4\cdot\psi_{a}^4)\Big)\right.\right. \nonumber \\
    & \left.\left.\qquad \left( \prod \limits_{c=1 \atop c\ne d}^4 \sum_{p_c=0}^{k_c} \binom{k_c}{p_c}\frac{\mathcal{M}_c^{p_c}}{(\alpha N)^{p_c}}\sum_{p_d=1}^{k_d} \binom{k_d}{p_d}\frac{p_d}{\alpha N}\frac{\mathcal{M}_d^{q+p_d-1}}{(\alpha N)^{p_d-1}} \right)\right] \right\rangle.
    \label{Loop-equation}
\end{align}
Let us now compute in detail the Leading Order (LO) and the Next to Leading Order (NLO) of the above Dyson-Schwinger Equation. Notice that the LO is of order $N^3$
 and it writes:
\begin{equation}
0=-\alpha\left\langle\text{Tr}\left(M^{(d)q}\right)\right\rangle+\frac{1}{\alpha}\sum \limits_{k \geq 1} \frac{1}{k}\sum \limits_{k_1+k_2+k_3+k_4=k}\left(-i\left(\frac{\lambda}{2}\right)^{\frac{1}{2}}\sigma^2\alpha\right)^k\frac{k!k_d}{k_1!k_2!k_3!k_4!}\left\langle\text{Tr}\left(M^{(d)q}\right)\right\rangle.
\label{LO-loop-eq}    
\end{equation}
This LO equation rewrites as:
\begin{equation}
    \alpha^2=\sum \limits_{k \geq 1} \frac{1}{k}\left(-i\left(\frac{\lambda}{2}\right)^{\frac{1}{2}}\sigma^2\alpha\right)^k \sum \limits_{k_1+k_2+k_3+k_4=k}\frac{k!k_d}{k_1!k_2!k_3!k_4!}.
\end{equation}
Let us now recall the
following identity: 
\begin{equation}
   \sum \limits_{k_1+\dots+k_D=k}\frac{k!}{k_1!\dots k_D!}\prod_{i=1}^D x_i^{k_i}=(x_1+\dots+x_D)^k.
\end{equation}
We now derive the above identity with respect to $x_d$; this leads to:
\begin{equation}
\label{multinomial_id_derivative}
    \sum \limits_{k_1+\dots+k_D=k}\frac{k!k_d}{k_1!\dots k_D!}\prod_{i \neq d}^D x_i^{k_i} x_d^{k_d-1}=k(x_1+\dots+x_D)^{k-1},
\end{equation}
Setting all the $x_i$'s equal to 1, we have:
\begin{equation}
    kD^{k-1}= \sum \limits_{k_1+\dots+k_D=k}\frac{k!k_d}{k_1!\dots k_D!}.
\end{equation}
Using the above result for $D=4$, the LO equation reduces to:
\begin{equation}
     \alpha^2=\sum \limits_{k \geq 1} \left(-i\left(\frac{\lambda}{2}\right)^{\frac{1}{2}}\sigma^2\alpha\right)^k 4^{k-1}=\frac{-i\alpha\sigma^2\sqrt{\lambda/2}}{1+4i\alpha\sigma^2\sqrt{\lambda/2}}.
\end{equation}
Finally, we can solve the LO of the Dyson-Schwinger equation and find the values of $\alpha$: 
\begin{equation}
    \alpha_{\pm}=\frac{-1\pm \sqrt{1+8\sigma^4\lambda}}{8i\sigma^2\sqrt{\lambda/2}}.
    \label{alpha}
\end{equation}
Notice that these are the same values of $\alpha_{\pm}$ found in \cite{Nguyen:2014mga} through a careful use of the saddle point method.

\medskip

Let us now evaluate the NLO of the Dyson-Schwinger equation. Collecting the terms of order $N^2$ in the Dyson-Schwinger equation \eqref{Loop-equation}, we get: 
\begin{align}
\label{NLO}
    & 0=\left\langle\sum_{i=0}^{q-1}\text{Tr}\left(M^{(d)i}\right)\text{Tr}\left(M^{(d){q-i-1}}\right)\right\rangle-N\left\langle\text{Tr}\left(M^{(d){q+1}}\right)\right\rangle+ \nonumber \\
    & +\frac{1}{\alpha^2} \left\langle\sum \limits_{k \geq 1} \frac{1}{k}\sum\limits_{k_1+k_2+k_3+k_4=k}\Big(-i\Big(\frac{\lambda}{2}\Big)^{\frac{1}{2}}\sigma^2\alpha\Big)^k\frac{k!}{k_1!k_2!k_3!k_4!}\sum_{c=1}^4 k_d(k_c-\delta_{cd})\text{Tr}\left(M^{(c)1-\delta_{cd}}\right)\text{Tr}\left(M^{(d)q+\delta_{cd}} \right)\right\rangle.
\end{align}
In order to solve the NLO equation, we first have to evaluate the third term in RHS of the equation above. 
in order to do this, we derive eq. \eqref{multinomial_id_derivative} with respect to $x_c$, and we sum over all flavours $c$. This leads to:
\begin{align}
    & \sum_{c=1}^D \frac{d}{d x_c} k (x_1+ \dots +x_D)^{k-1}=D k(k-1)(x_1+ \dots +x_D)^{k-2}= \nonumber \\
    & \sum_{c=1}^D \sum \limits_{k_1+\dots+k_D=k}\frac{k!k_d(k_c-\delta_{cd})}{k_1!\dots k_D!}\prod_{i \neq c,d}^D x_i^{k_i} x^{k_c-\delta_{cd}} x_d^{k_d-1-\delta_{cd}}.
\end{align}
As above, let us set all the $x_i$'s equal to 1, and insert the resulting identity for $D=4$ in the NLO Dyson-Schwinger equation \eqref{NLO}. We get:
\begin{align}
    & 0=\left\langle\sum_{i=0}^{q-1}\text{Tr}\left(M^{(d)i}\right)\text{Tr}\left(M^{(d){q-i-1}}\right)\right\rangle-N\left\langle\text{Tr}\left(M^{(d){q+1}}\right)\right\rangle+ \nonumber \\
    & +\frac{1}{\alpha^2} \sum \limits_{k \geq 1} \Big(-i\Big(\frac{\lambda}{2}\Big)^{\frac{1}{2}}\sigma^2\alpha\Big)^k (k-1)4^{k-1} \left\langle\sum_{c=1}^4 \text{Tr}\left(M^{(c)1-\delta_{cd}}\right)\text{Tr}\left(M^{(d)q+\delta_{cd}} \right)\right\rangle.
\end{align}
Using eq. \eqref{alpha} the NLO term of the Dyson-Schwinger equation reduces to
\begin{equation}
    0=\left\langle\sum_{i=0}^{q-1}\text{Tr}\left(M^{(d)i}\right)\text{Tr}\left(M^{(d){q-i-1}}\right)\right\rangle+(\alpha^2-1)N\left\langle\text{Tr}\left(M^{(d)q+1}\right)\right\rangle+\alpha^2\left\langle\sum_{c=1 \atop c\neq d}^4 \text{Tr}\left(M^{(c)}\right)\text{Tr}\left(M^{(d)q}\right)\right\rangle.
\end{equation}
Notice that, also in the NLO term of the Dyson-Schwinger equation, we recovered exactly the same result of \cite{Nguyen:2014mga}. Even if we considered a non-Gaussian distribution (a Gaussian term plus a quartic pillow term potential), the first orders of the Dyson-Schwinger equation are the same as in the  Gaussian case.

\subsection{Gross-Rosenhaus SYK generalization}

Let us first consider the formula \eqref{Int.Field_Action_Gross-Ros} for the field $M^{(c)}$ and perform the following change of variables: 
\begin{equation}
    M^{(c)} \rightarrow \alpha \mathbb{1} + \frac{N}{\prod_{a=1}^f N_a^{\frac{q_a}{2}}} M^{(c)}.
\end{equation}
The action \eqref{Int.Field_Action_Gross-Ros} thus rewrites:
\begin{align}
    &\mathcal{S}[M^{(c)}]
    =\frac{1}{2}N  \sum_{c=1}^q \text{Tr}\left( (M^{(c)})^2 \right) + \alpha \prod_{a=1}^f N_a^{\frac{q_a}{2}} \sum_{c=1}^q \text{Tr}\left( M^{(c)}) \right) \nonumber \\
    & - \frac{1}{2}\text{Tr}\left[ \sum_{k \ge 1} \left( -i \alpha \sqrt{2\lambda} \sigma^2 \prod_{a=1}^f (q_a-1)! \right)^k \sum_{\sum_i^q k_i=k}\binom{k}{k_1, \dots ,k_q} \hat{A}_k \prod_{c=1}^q \sum_{p_c=0}^{k_c} \binom{k_c}{p_c} \left( \frac{N}{\alpha \prod_{a=1}^f N_a^{\frac{q_a}{2}}}  \right)^{p_c} \mathcal{M}_c^{p_c} \right] 
\end{align}
The Dyson-Schwinger equation for the intermediate field writes:
\begin{equation}
    0=\sum_{ij}\int \prod_{c=1}^q \mathrm{d}M^{(c)} \frac{\partial}{\partial M_{ij}^{(d)}} \left[ \left( M^{(d)} \right)_{ij}^{h} e^{-\mathcal{S}[M]} \right] ,
\end{equation}
This leads to:
\begin{align}
    & 0=\left\langle\sum_{i=0}^{h-1}\text{Tr}\left(M^{(d)i}\right)\text{Tr}\left(M^{(d){h-i-1}}\right)\right\rangle-N \left\langle\text{Tr}\left(M^{(d){h+1}}\right)\right\rangle-\alpha \prod_{a=1}^f N_a^{\frac{q_a}{2}} \left\langle\text{Tr}\left(M^{(d)h}\right)\right\rangle+ \nonumber \\
    & +\left\langle \text{Tr} \left[ \sum_{k \ge 1} \left( -i \alpha \sqrt{2\lambda} \sigma^2 \prod_{a=1}^f (q_a-1)! \right)^k \sum_{\sum_i^q k_i=k}\binom{k}{k_1, \dots ,k_q} \hat{A}_k \right. \right. \nonumber \\
    & \left.\left. \left( \prod_{c=1 \atop c \neq d}^q \sum_{p_c=0}^{k_c} \binom{k_c}{p_c} \left( \frac{N}{\alpha \prod_{a=1}^f N_a^{\frac{q_a}{2}}}  \right)^{p_c} \mathcal{M}_c^{p_c}  \sum_{p_d=1}^{k_d} \binom{k_d}{p_d} p_d \left(\frac{N}{\alpha \prod_{a=1}^f N_a^{\frac{q_a}{2}}}  \right)^{p_d} \mathcal{M}_d^{h+p_d-1}   \right) \right] \right\rangle.
    \label{Loop-equation2}
\end{align}
Let us define $\kappa_a = \frac{N_a}{N}$, so that the LO of the Dyson-Schwinger equation in the large $N$ limit is of the order of $N^{\frac{q_a+2}{2}}$.
The LO contribution thus writes:
\begin{align}
&0=-\alpha N^{\frac{q_a}{2}+1} \prod_{a=1}^f \kappa_a^{\frac{q_a}{2}}\left\langle\text{Tr}\left(M^{(d)h}\right)\right\rangle \\
&+\frac{N^{\frac{q_a}{2}+1} \prod_{a=1}^f \kappa_a^{\frac{q_a}{2}}}{\alpha\kappa_d}\sum \limits_{k \geq 1} \frac{1}{k} \sum_{\sum_i^q k_i=k}\binom{k}{k_1, \dots ,k_q}\left(-i\left(\frac{\lambda}{2}\right)^{\frac{1}{2}}\sigma^2\alpha\prod_{a=1}^f (q_a-1)! \right)^kk_d\left\langle\text{Tr}\left(M^{(d)h}\right)\right\rangle. \nonumber
\end{align}
which becomes
\begin{equation}
     \alpha^2=\frac{1}{\kappa_d}\sum \limits_{k \geq 1} \left(-i\left(\frac{\lambda}{2}\right)^{\frac{1}{2}}\sigma^2\alpha\prod_{a=1}^f (q_a-1)! \right)^k q^{k-1}=\frac{1}{\kappa_d}\frac{-i\alpha\sigma^2\sqrt{\lambda/2}\prod_{a=1}^f (q_a-1)! }{1+qi\alpha\sigma^2\sqrt{\lambda/2}\prod_{a=1}^f (q_a-1)! },
\end{equation}
with
\begin{equation}
    \alpha_{\pm}=\frac{-1\pm \sqrt{1+2q\sigma^4\frac{\lambda}{\kappa_d}(\prod_{a=1}^f (q_a-1)!)^2}}{2iq\sigma^2\sqrt{\lambda/2}\prod_{a=1}^f (q_a-1)!}.
\end{equation}
We can note that the saddle point is parametrized by $\kappa_d$.

\medskip

The NLO part of the Dyson-Schwinger equations writes:
\begin{align}
     &0=\left\langle\sum_{i=0}^{h-1}\text{Tr}\left(M^{(d)i}\right)\text{Tr}\left(M^{(d){h-i-1}}\right)\right\rangle-N\left\langle\text{Tr}\left(M^{(d){h+1}}\right)\right\rangle \nonumber \\
    &+\sum \limits_{k \geq 1} \frac{1}{k} \sum_{\sum_i^q k_i=k}\binom{k}{k_1, \dots ,k_q}\Big(-i\Big(\frac{\lambda}{2}\Big)^{\frac{1}{2}}\sigma^2\alpha\prod_{a=1}^f (q_a-1)! \Big)^k \sum_{c=1}^q \frac{k_d(k_c-\delta_{cd})}{\kappa_d\kappa_c\alpha^2}\left\langle\text{Tr}\left(M^{(c)1-\delta_{cd}}\right)\text{Tr}\left(M^{(d)h+\delta_{cd}} \right)\right\rangle.
\end{align}
This can be rewritten as:
\begin{align}
     &0=\left\langle\sum_{i=0}^{h-1}\text{Tr}\left(M^{(d)i}\right)\text{Tr}\left(M^{(d){h-i-1}}\right)\right\rangle-N\left\langle\text{Tr}\left(M^{(d){h+1}}\right)\right\rangle \nonumber \\
    &+\sum \limits_{k \geq 1}\Big(-i\Big(\frac{\lambda}{2}\Big)^{\frac{1}{2}}\sigma^2\alpha\prod_{a=1}^f (q_a-1)! \Big)^k(k-1)q^{k-1}  \sum_{c=1}^q\frac{1}{\kappa_d\kappa_c\alpha^2}\left\langle\text{Tr}\left(M^{(c)1-\delta_{cd}}\right)\text{Tr}\left(M^{(d)h+\delta_{cd}} \right)\right\rangle,
\end{align}
Finally, we get the following equation:
\begin{align}
     &0=\left\langle\sum_{i=0}^{h-1}\text{Tr}\left(M^{(d)i}\right)\text{Tr}\left(M^{(d){h-i-1}}\right)\right\rangle-N\left\langle\text{Tr}\left(M^{(d){h+1}}\right)\right\rangle + q\alpha^2\sum_{c=1}^q\frac{\kappa_d}{\kappa_c}\left\langle\text{Tr}\left(M^{(c)1-\delta_{cd}}\right)\text{Tr}\left(M^{(d)h+\delta_{cd}} \right)\right\rangle.
\end{align}

\bibliographystyle{unsrt}
\bibliography{biblio}

\end{document}